\documentclass[letterpaper, 12pt]{article}
\pdfoutput=1
\usepackage[margin=0.75in]{geometry}
\usepackage{graphicx}
\usepackage{amsmath}
\usepackage{microtype}
\usepackage{mathpazo}
\usepackage{tabularx}
\usepackage{multirow}
\usepackage{booktabs}
\usepackage{longtable}

\usepackage[table]{xcolor}
\definecolor{lightgray}{gray}{0.9}
\definecolor{darkgray}{gray}{0.8}

\usepackage{hyperref}
\title{Deciphering the regulatory genome of \textit{Escherichia
coli}, one hundred promoters at a time}

\date{\vspace{-9ex}}

\begin{document}

\maketitle

\noindent William T. Ireland\textsuperscript{1}, 
Suzannah M. Beeler\textsuperscript{2}
Emanuel Flores-Bautista\textsuperscript{2},
Nathan M. Belliveau\textsuperscript{2,\textdagger},
Michael J. Sweredoski\textsuperscript{3},
Annie Moradian\textsuperscript{3},
Justin B. Kinney\textsuperscript{4},
Rob Phillips\textsuperscript{1,2,5,*} \\ 

\noindent \textsuperscript{1} Department of Physics, California Institute of Technology, Pasadena, CA 91125 \\
\textsuperscript{2} Division of Biology and Biological Engineering, California Institute of Technology, Pasadena, CA 91125 \\
\textsuperscript{3} Proteome Exploration Laboratory, Beckman Institute, California Institute of Technology, Pasadena, CA 91125 \\ 
\textsuperscript{4} Simons Center for Quantitative Biology, Cold Spring Harbor Laboratory, Cold Spring Harbor, NY 11724 \\
\textsuperscript{5} Department of Applied Physics, California Institute of Technology, Pasadena, CA 91125 \\
\textsuperscript{\textdagger} Present address: {Howard Hughes Medical Institute and Department of Biology, University of Washington, Seattle, WA 98195 \\
\textsuperscript{*} Corresponding author: phillips@pboc.caltech.edu


\begin{abstract}
    Advances in DNA sequencing have revolutionized our ability to read genomes. However, even in the most well-studied of organisms, the bacterium  {\it Escherichia coli}, for $\approx$ 65\% of the promoters we remain completely ignorant of their regulation.  Until we have cracked this
regulatory Rosetta Stone, efforts to read and write genomes will remain 
haphazard. We introduce a new method (Reg-Seq) linking a massively-parallel reporter assay and mass spectrometry to produce a base pair resolution dissection of more than 100 promoters in {\it E. coli}  in 12 different growth conditions. 
First, we show that our method recapitulates regulatory information from known sequences.  Then,
we examine the regulatory architectures for more than 80 promoters in
the {\it E. coli} genome which previously had no known regulation.  In many cases, we also identify which transcription factors mediate their regulation.
The method introduced here clears a
path for fully characterizing the regulatory genome of model organisms, with the potential of moving on to an
array of other microbes of ecological and medical relevance.
\end{abstract}

\section{Introduction} 

DNA sequencing is as important to biology as the telescope is to astronomy. We are now living in the age of genomics, where DNA sequencing has become cheap and routine. However, despite these incredible advances, how all of this genomic information is regulated and deployed remains largely enigmatic. Organisms must respond to their environments through regulation of genes. Genomic methods often provide a ``parts'' list but often leave us uncertain about how those parts are used creatively and constructively in space and time.  Yet, we know that promoters apply all-important dynamic logical operations that control when and where genetic information is accessed.   In this paper, we demonstrate how we can infer the logical and regulatory interactions that control bacterial decision making by tapping into the  power of DNA sequencing as a biophysical tool. The method introduced here provides a framework for solving the problem of deciphering the regulatory genome by connecting perturbation and response, mapping information flow from individual nucleotides in a promoter sequence to downstream gene expression,  determining how much information each promoter base pair carries about the level of gene expression.\\

The advent of RNA-Seq \cite{Mortazavi2008} launched a new era in which sequencing could be used as an experimental read-out of the biophysically interesting counts of mRNA, rather than simply as a tool for collecting ever more complete organismal genomes. 
The slew of `X'-Seq technologies that are available continues to expand at a dizzying pace, each serving their own creative and insightful role: RNA-Seq, ChIP-Seq, Tn-Seq, SELEX, 5C, etc. \cite{Stuart2019}.
In contrast to whole genome screening sequencing approaches, such as Tn-Seq \cite{Goodall2017} and ChIP-Seq \cite{Gao2018} which give a coarse-grained view of gene essentiality and regulation respectively, another class of experiments known as massively-parallel reporter assays
(MPRA) has been used to study gene expression in a variety of contexts \cite{Patwardhan2009, Kinney2010, Sharon2012,Patwardhan2012,Melnikov2012, Kwasnieski2012,Fulco2019,Kinney2019}.
One elegant study relevant to the bacterial case of interest here by \cite{Kosuri2013}  screened more than $10^4$ combinations of promoter and ribosome binding sites (RBS). Even more recently, they have utilized MPRAs in sophisticated ways to search for regulated genes across the genome \cite{Urtecho2019, Urtecho2020}, in a way we see as being complementary to our own. While their approach yields a coarse-grained view of where regulation may be occurring, our approach yields a base-pair-by-base-pair view of how exactly that regulation is being enacted. \\

One of the most exciting X-Seq tools based on MPRAs with broad biophysical reach is the Sort-Seq approach developed by \cite{Kinney2010}. 
Sort-Seq uses fluorescence activated cell sorting (FACS) based on changes in the fluorescence due to mutated promoters to identify the specific locations of transcription factor binding in the genome.
Importantly, it also provides a readout of how promoter sequences  control the level of gene expression with single base-pair resolution.
The results of such a massively-parallel reporter assay make it possible to build a biophysical model of gene regulation to uncover how previously uncharacterized promoters are regulated.   
 In particular, high-resolution studies like those described here yield quantitative predictions about promoter organization and protein-DNA interactions as described by energy matrices~\cite{Kinney2010}.  This allows us to employ the tools of statistical physics to describe the input-output properties of each of these promoters which can be explored much further with in-depth experimental dissection like those done by~\cite{Razo-Mejia2018} and \cite{Chure2019} and summarized in ~\cite{Phillips2019}.  In this sense, the Sort-Seq approach can provide a quantitative framework to not only discover and quantitatively dissect regulatory interactions at the promoter level, but also provides an interpretable scheme to design genetic circuits with a desired expression output \cite{Barnes2019}.
 \\

Earlier work from \cite{Belliveau2018} illustrated how Sort-Seq,  used in conjunction with mass spectrometry can be used to identify which transcription factors bind to a given binding site,  thus
enabling the mechanistic dissection of promoters which  previously had no regulatory annotation. 
 However, a crucial drawback of the approach of \cite{Belliveau2018} is that while it is high-throughput at the level of a single gene and the number of promoter variants it accesses, it was unable to readily tackle multiple genes at once, still leaving much of the unannotated genome untouched.  Given that even in one of biology's best understood organisms, the bacterium  \textit{Escherichia coli}, for more than 65\% of its genes, we remain completely ignorant of how those genes are regulated \cite{Santos_Zavaleta2019,Belliveau2018}. If we hope to some day have a complete base pair resolution mapping of how genetic sequences relate to biological function, we must first be able to do so for the promoters of this ``simple" organism. \\

What has been missing in uncovering the regulatory genome in organisms of
 all kinds  is a  large scale method for inferring genomic logic and regulation.    Here we replace the low-throughput fluorescence-based Sort-Seq approach with a scalable RNA-Seq based approach that makes it possible to attack multiple promoters at once, setting the stage for the possibility of, to first approximation, uncovering the entirety of the regulatory genome. Accordingly, we refer to the entirety of our approach (MPRA, information footprints and energy matrices, mass spectrometry for transcription factor identification) as Reg-Seq, which we employ here on over one hundred promoters. 
The concept of  MPRA methods is to perturb promoter regions by  mutating them and then using sequencing to read out both perturbation and the resulting gene expression \cite{Patwardhan2009, Kinney2010, Sharon2012,Patwardhan2012,Melnikov2012, Kwasnieski2012,Fulco2019,Kinney2019}. We generate a broad diversity of promoter sequences  for each promoter of interest and use mutual information as a metric to measure information flow from that distribution of sequences to gene expression. 
Thus, Reg-Seq is able to collect causal information about candidate regulatory sequences that is then complemented by mass spectrometry which allows us to find which transcription factors
mediate the action of those newly discovered  candidate regulatory sequences. Hence, Reg-Seq solves the causal problem of linking DNA sequence to regulatory logic and information flow.\\

To demonstrate our ability to scale up Sort-Seq with the sequencing based Reg-Seq protocol, we report here our results for 113 {\it E. coli} genes, whose regulatory architectures (i.e. gene-by-gene distributions of transcription factor (TF) binding sites and identities of TFs that bind those sites)
were determined in parallel. 
 By taking the Sort-Seq approach from a gene-by-gene method to a more whole-genome approach, we can begin to piece together not just how individual promoters are regulated, but also
the nature of gene-gene interactions by revealing  how certain transcription factors serve to regulate multiple genes at once. This approach has the benefits of a high-throughput assay while sacrificing little of the resolution afforded by the previous  gene-by-gene approach, allowing us to uncover a large swath of the \textit{E. coli} regulome, with base-pair resolution, in one set of experiments. \\

The organization of the remainder of the paper is as follows. In the Results section, we provide a global view of the discoveries we made in our exploration of more than 100 promoters in {\it E. coli} using Reg-Seq. These results are described in summary form in the paper itself, with a full online version of the results (\url{www.rpgroup.caltech.edu/RNAseq_SortSeq/interactive_a}) showing how different growth conditions elicit different regulatory responses. This section also  follows the overarching view of our results by examining several biological stories that emerge from our data and serve as case studies in what has been revealed in our efforts to uncover the regulatory genome. The Discussion section summarizes the method and the current round of discoveries it has afforded with an eye to future applications  to further elucidate the {\it E. coli} genome and opening up the quantitative dissection of other non-model organisms. Lastly, in the Methods section and fleshed out further in the Supplementary Information, we describe our methodology and benchmark it against our own earlier Sort-Seq experiments to show that using RNA-Seq as a readout of the expression of mutated promoters is equally reliable as the fluorescence-based approach. \\

\section{Results} \label{results}

\subsection{Selection of genes and methodology}

As shown in Figure~\ref{fig:genome}, we have considered more than 100 genes from across the {\it E. coli} genome. 
Our choices were based on a number of factors (see Sections 1.1 and 1.2 of the SI for more details); namely, we wanted a subset of genes that served 
as a ``gold standard'' for which the hard work of generations of molecular biologists have yielded deep insights into their regulation. The set includes \textit{lacZYA}, \textit{znuCB}, \textit{znuA}, \textit{ompR}, \textit{araC}, \textit{marR}, \textit{relBE}, \textit{dgoR}, \textit{dicC}, \textit{ftsK}, \textit{xylA}, \textit{xylF}, \textit{dpiBA}, \textit{rspA}, \textit{dicA}, and \textit{araAB}. By using Reg-Seq on these genes we were able to demonstrate
that this method recovers
 not only what was already known about binding sites and transcription factors for well-characterized promoters, but also  whether there are any important differences between the results of the methods presented here and the previous generation of experiments based on fluorescence and cell-sorting as a readout of gene expression. These promoters of known regulatory architecture are complemented by an array of previously uncharacterized genes that we selected in part using data from a recent proteomic study, in which mass spectrometry was used to measure the copy
 number of different proteins in 22 distinct growth conditions~\cite{Schmidt2015}.
 We selected genes that exhibited a wide variation in their copy number over the different growth conditions considered, reasoning that differential expression across growth conditions implies that those genes are under regulatory control. \\

\begin{figure}
\centering
\includegraphics[width=6.5truein]{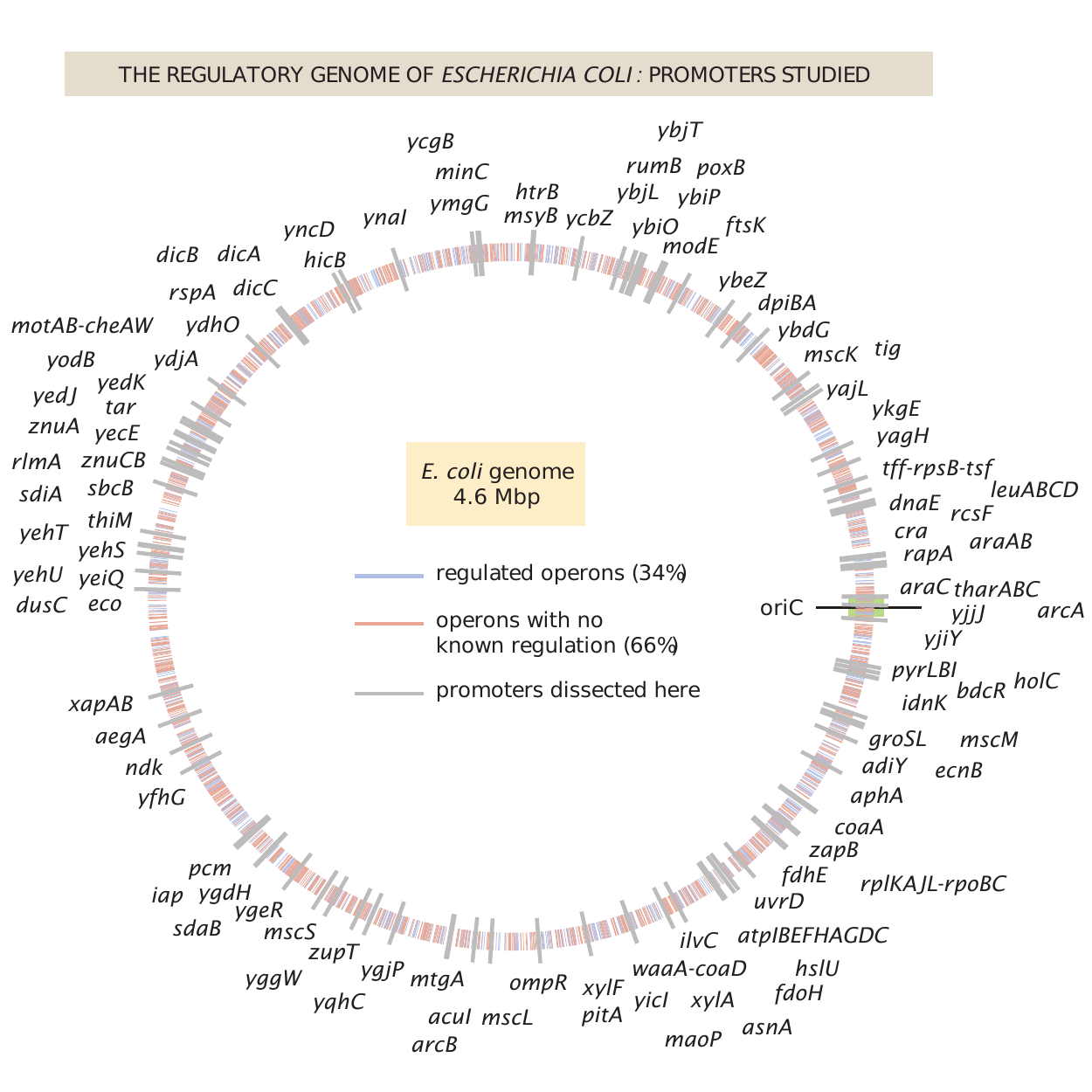}
\caption{The {\it E. coli} regulatory genome.  Illustration of the current ignorance with respect to how genes are regulated in {\it E. coli}, with genes with previously annotated regulation (as reported on RegulonDB~\cite{Gama-Castro2016}) denoted with blue ticks and genes with no previously annotated regulation denoted with red ticks. The 113 genes explored in this study are labeled in gray.}
\label{fig:genome}
\end{figure}

\begin{figure}
\centering
\includegraphics[width=4.75truein]{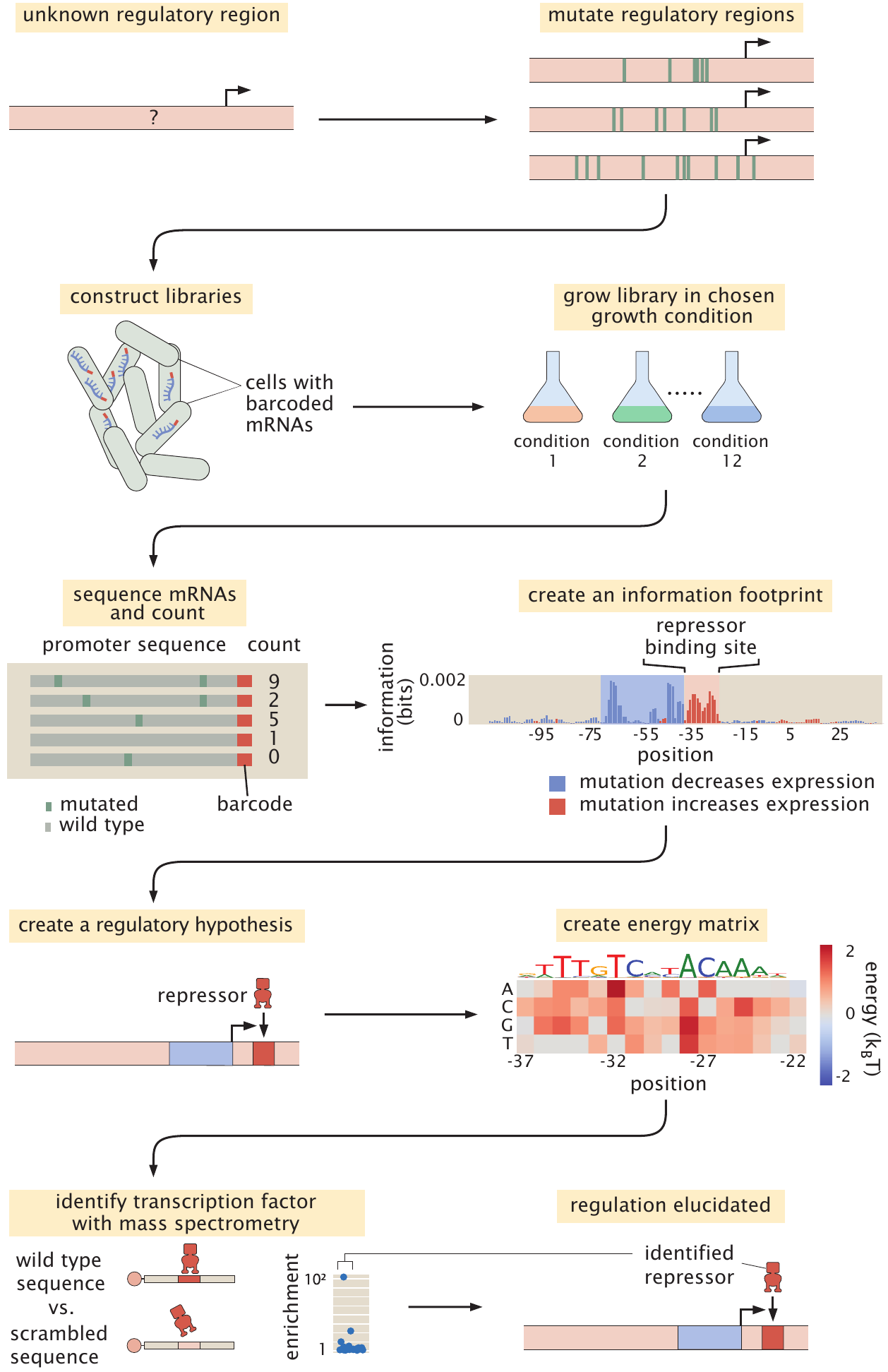}
\caption{The Reg-Seq procedure used to determine how a given promoter is regulated. This process is as follows: After constructing a promoter library driving expression of a randomized barcode (an average of 5 for each promoter), RNA-Seq is conducted to determine frequency of these mRNA barcodes across different growth conditions. By computing the mutual information between DNA sequence and mRNA barcode counts for each base pair in the promoter region, an "information footprint" is constructed yielding a regulatory hypothesis for the putative binding sites. Energy matrices, which describe the effect any given mutation has on DNA binding energy, and sequence logos are inferred for the putative transcription factor binding sites. Next, we identify which transcription factor preferentially binds to the putative binding site via DNA affinity chromatography followed by mass spectrometry. Finally, this procedure culminates in a coarse-grained cartoon-level view of our regulatory hypothesis for how this given promoter is regulated.}
\label{fig:procedure}
\end{figure}

As noted in the introduction, the original formulation of Reg-Seq termed
Sort-Seq was based on the use of fluorescence activated cell sorting one gene at a time as a way to uncover putative binding sites for previously uncharacterized promoters \cite{Belliveau2018}.  As a result, as shown in Figure~\ref{fig:procedure} we have formulated a second generation version that permits a high-throughput interrogation of the genome. 
A comparison between the Sort-Seq and Reg-Seq approaches for the same genes is shown in Supplemental Figure S1.
In the Reg-Seq approach, for each promoter interrogated, we generate a library of mutated variants and design each variant to express an mRNA with a unique sequence barcode. By counting the frequency of each expressed barcode using RNA-Seq, we can assess
 the differential expression from our promoter of interest based on the base-pair-by-base-pair sequence of its promoter. Using the mutual information between mRNA counts and sequences, we develop an information footprint that reveals the importance of different bases in the promoter region to the overall level of expression. We locate potential transcription factor binding regions by looking for clusters of base pairs that have a significant effect on gene expression.
  Further details on how potential binding sites are identified are found in the Methods section.  Blue regions of the histogram shown in
 the information footprints of Figure~\ref{fig:procedure} correspond to hypothesized activating sequences and red regions of the histogram correspond to hypothesized repressing sequences. With the information footprint in hand, we can then determine energy matrices and sequence logos (described in the next section). Given putative binding sites, we construct oligonucleotides that serve as fishing hooks to fish out the transcription factors that bind to those putative binding sites
  using DNA-affinity chromatography and mass spectrometry~\cite{Mittler2009}. Given all of this information, we can then formulate a schematized view of the newly discovered regulatory architecture of the previously uncharacterized promoter. For the case schematized in Figure~\ref{fig:procedure}, the experimental pipeline yields a complete picture of a simple repression architecture (i.e. a gene regulated by a single binding site for a repressor). \\

\subsection{Visual tools for data presentation}

Throughout our investigation of the more than 100 genes explored in this study, we repeatedly relied on several key approaches to help make sense of the immense amount of data generated in
these experiments. As these different approaches to viewing the results will appear repeatedly throughout the paper, here we familiarize the reader with  five graphical representations referred to respectively as information footprints, energy matrices, sequence logos,  mass spectrometry enrichment plots and regulatory cartoons, which taken all together provide a quantitative description of previously uncharacterized promoters. \\

{\it Information footprints:} From our mutagenized libraries of promoter regions, we can build up a base-pair-by-base-pair graphical understanding of how the promoter sequence relates to level of gene expression  in the form of the information footprint shown in the middle of
Figure~\ref{fig:procedure}. In this plot, the bar above each base pair position represents how large of an effect mutations at this location have on the level of gene expression. Specifically, the quantity plotted   is the mutual information $I_b$ at base pair $b$  between mutation of a base pair at that position and the level of expression.  In mathematical terms, the mutual information
measures how much the joint probability $p(m,\mu)$ differs from the product of
the probabilities $p_{mut}(m)p_{expr}(\mu)$ which would be produced if mutation and gene expression level
were independent.  Formally, the mutual information between having a mutation at position $b$ and
level of expression is given by
\begin{equation}
    I_b =  \sum_{m=0}^1  \sum_{\mu=0}^1 p(m,\mu)\log_2\left(\frac{p(m,\mu)}{p_{mut}(m)p_{expr}(\mu)}\right).
    \label{eqn:MI}
\end{equation}
\noindent Note that both $m$ and $\mu$ are binary variables that characterize the mutational state of the base of interest and the level of expression, respectively.  Specifically, 
 $m$ can take  the values
\begin{equation}
    m=
    \begin{cases}
      0, & \text{if}\ b \mbox{ is a mutated base} \\
      1, & \text{if}\ b \mbox{ is a wild-type base.}
    \end{cases}
  \end{equation}
  and $\mu$ can take on values 
   \begin{equation}
    \mu=
    \begin{cases}
      0, & \text{for sequencing reads from the DNA library}  \\
      1, & \text{for sequencing reads originating from mRNA,}
    \end{cases}
  \end{equation}
  \noindent  where  both $m$ and $\mu$ are index variables that tell us whether the base has been mutated and if so, how likely that the read at that position will correspond to an mRNA, reflecting gene expression or a promoter, reflecting a member of the library.  The higher the ratio of mRNA to DNA reads at a given base position, the higher the expression.  $p_{mut}(m)$ in equation~\ref{eqn:MI} refers to the probability that a given sequencing read will be from a mutated base. $p_{expr}(\mu)$ is a normalizing factor that gives the ratio of the number of DNA or mRNA sequencing counts to total number of counts.   \\

Furthermore, we color the bars based on whether mutations at this location lowered gene expression on average  (in blue, indicating an activating role) or increased gene expression (in red, indicating a repressing role). Within these footprints, we look for regions of approximately 10 to 20 contiguous base pairs which impact gene expression similarly (either increasing or decreasing), as these regions implicate the influence of a transcription factor binding site. In this experiment, we targeted the regulatory regions based on a guess of where a transcription start site (TSS) will be, based on experimentally confirmed sites contained in regulonDB \cite{Santos_Zavaleta2019}, a 5' RACE experiment 
~\cite{mendoza-vargas_genome-wide_2009}, or by targeting small intergenic regions. After completing the Reg-Seq experiment, we note that many of the presumed TSS sites are not in the locations assumed, the promoters have multiple active  RNA polymerase (RNAP) sites and TSS, or the primary TSS shifts with growth condition. To simplify the data presentation, the '0' base pair in all information footprints is set to the originally assumed base pair for the primary TSS, rather than one of the TSS that was found in the experiment. 
As can be seen throughout the paper (see Figure~\ref{fig:architectures} for several examples of each of the main types of regulatory architectures) and the online resource, we present such information footprints for every promoter we have considered, with one such information footprint for every growth condition.\\

{\it Energy matrices:} Focusing on an individual putative transcription factor binding site as revealed in
the information footprint, we are interested in a more fine-grained, quantitative understanding of how the underlying protein-DNA interaction is determined. An energy matrix displays this information using a heat map format, where each column is a position in the putative binding site and each row displays the effect on binding that results from mutating to that given nucleotide (given as a change in the DNA-TF interaction energy upon mutation) \cite{Kinney2010,Berg1987,Stormo1998}. These energy matrices are scaled such that the wild type sequence is colored in white, 
mutations that improve binding are shown in blue, and mutations that weaken binding are shown in red. These energy matrices encode a full quantitative picture for how we expect sequence to relate to binding for a given transcription factor, such that we can provide a prediction for the binding energy of every possible binding site sequence as
\begin{equation}
\mbox{binding energy}= \sum_{i=1}^N \varepsilon_i,
\end{equation}
where the energy matrix is predicated on an assumption of a linear binding model in which each base within
the binding site region contributes a specific value ($\varepsilon_i$ for the $i^{th}$ base in
the sequence) to the total binding energy. Energy matrices are either given in A.U. (arbitrary units), or if the gene has a simple repression or activation architecture with a single RNA polymerase (RNAP) site, are assigned k$_B$T energy units following the procedure in \cite{Kinney2010} and validated on the \textit{lac} operon in \cite{Barnes2019}. \\

{\it Sequence logos:} From an energy matrix, we can also represent a preferred transcription factor binding site with the use of the letters corresponding to the four possible nucleotides, as is often done with position weight matrices \cite{schneider_sequence_1990}. In these sequence logos, the size of the letters corresponds to how strong the preference is for that given nucleotide at that given position, which can be directly computed from the energy matrix. This method of visualizing the information contained within the energy matrix is more easily digested and allows for quick comparison among various binding sites. \\

{\it Mass spectrometry enrichment plots:} As the final piece of our experimental pipeline, we wish to determine the identity of the transcription factor we suspect is binding to our putative binding site that is represented in the energy matrix and sequence logo. While the details of the DNA  affinity chromatography and mass spectrometry can be found in the methods, the results of these experiments are displayed in enrichment plots such as is shown in
the bottom panel of Figure~\ref{fig:procedure}. In these plots, the relative abundance of each protein bound to our site of interest is quantified relative to a scrambled control sequence. The putative transcription factor is the one we find to be highly enriched compared to all other DNA binding proteins.\\

{\it Regulatory cartoons:} The ultimate result of all these detailed base-pair-by-base-pair resolution experiments yields a cartoon model of how we think the given promoter is being regulated. A complete set of cartoons for all the architectures considered in our
study is presented in Supplemental Figure S6. While the cartoon serves as a convenient visual way to summarize our results, it's important to remember that these cartoons are a shorthand representation of all the data in the four quantitative measures described above and are in fact backed by quantitative predictions of how we expect the system to behave which can be tested experimentally. Throughout this paper  we use consistent iconography to illustrate the regulatory architecture of promoters, with activators and their binding sites in green, repressors in red, and RNAP in blue. \\

\subsection{Newly discovered {\it E. coli} regulatory architectures}

Figure~\ref{fig:counts_summary} (and Tables~\ref{tab:allprom} and ~\ref{tab:allgenes}) provides a summary of the discoveries made in the work done here using our next generation Reg-Seq approach. Figure~\ref{fig:counts_summary}(A) provides a shorthand notation that conveniently characterizes the different kinds of regulatory architectures found in bacteria. In previous work~\cite{Rydenfelt2014-2}, we have explored the entirety of what is known about the regulatory genome of {\it E. coli}, revealing that the most common motif is the (0,0) constitutive architecture, though we hypothesized that this is not a statement about the facts of the {\it E. coli} genome, but rather a reflection of our collective regulatory ignorance in the sense that we suspect that with further investigation, many of these apparent constitutive architectures will be found to be regulated under the right environmental conditions.  The two most common regulatory architectures that emerged from our previous database survey are the (0,1) and (1,0) architectures, the simple repression motif and the simple activation motif, respectively. It is interesting to consider that the (0,1) architecture is in  fact the repressor-operator model originally introduced in the early 1960s by Jacob and Monod as the concept of gene regulation emerged~\cite{jacob_regulation_1961}.   Now we see retrospectively the far-reaching importance of that architecture across the {\it E. coli} genome. \\

For the 113 genes we considered, Figure~\ref{fig:counts_summary}(B) summarizes  the number of
simple repression $(0,1)$ architectures discovered, the number of simple activation $(1,0)$ architectures discovered and so on.  A comparison of the frequency of the different architectures found in our
study to the frequencies of all the known architectures in the RegulonDB database is provided
in Supplemental Figure S7.  Tables~\ref{tab:allprom} and ~\ref{tab:allgenes} provide a more detailed view of our results.  As seen in Table~\ref{tab:allprom}, of the 113 genes we considered, 32 of them revealed 
 no signature of any transcription
factor binding sites and they are labeled as $(0,0)$. The simple repression architecture  $(0,1)$ was found 26 times, the simple activation architecture $(1,0)$ was found 13 times, and more complex architectures featuring multiple binding sites (e.g. $(1,1)$, $(0,2)$, $(2,0)$, etc.)  were revealed as well.   Further, for 18 of the genes that we
label ``inactive'',  Reg-Seq didn't even reveal an RNAP binding site. The lack of observable RNAP site could be because the proper growth condition to get high levels of expression was not used, or because the mutation window chosen for the gene does not capture a highly transcribing TSS. 
The tables also include our set of 16 ``gold standard'' genes for which previous work has resulted in a knowledge (sometimes only partial) of their regulatory architectures. We find that our method recovers the regulatory elements of these gold standard cases fully in 12 out of 16 cases, and the majority of regulatory elements in 2 of the remainder. Overall the performance of
Reg-Seq in these gold-standard cases (for more details see Supplemental Figure S2) builds confidence in the approach. Further, the failure modes inform us of the blind spots of Reg-Seq. For example, we find it challenging to observe weaker binding sites when multiple strong binding sites are also present such as in the \textit{marRAB} operon. Additionally the method will fail when there is no active TSS in the mutation window, as occurred in the case of \textit{dicA}. Further details on the comparison to gold standard genes can be found in SI Section 2.2.\\

\begin{figure}[!h]
\centering
\includegraphics[width=6.5truein]{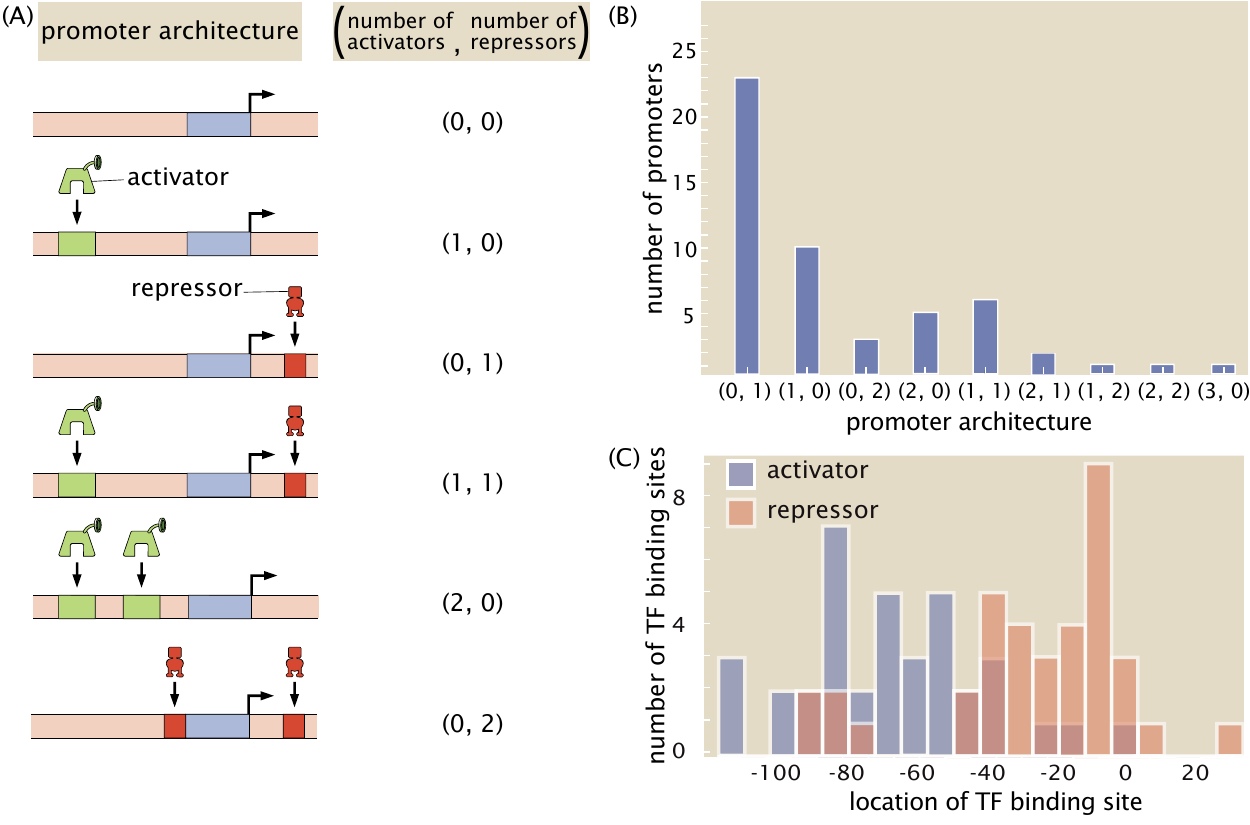}
\caption{A summary of regulatory architectures discovered in this study. (A) The cartoons display a representative example of each type of architecture, along with the corresponding shorthand notation. (B) Counts of the different regulatory architectures discovered in this study.  Only those promoters where at least one new binding site was discovered are included in this figure. If one repressor was newly discovered and two activators were previously known, then the architecture is still counted as a (2,1) architecture. (C) Distribution of positions of binding sites discovered in this study for activators and repressors. Only newly discovered binding sites are included in this figure. The position of the TF binding sites are calculated relative to the estimated TSS location, which is based on the location of the associated RNAP site.}
\label{fig:counts_summary}
\end{figure}

We observe that the most common motif to emerge from our work is the simple repression motif. Another relevant regulatory statistic is shown in Figure~\ref{fig:counts_summary}(C)  where we see the distribution of binding site positions. Our own experience in the use of different quantitative modeling approaches to consider transcriptional regulation reveal that, for now, we remain largely ignorant of how to account for transcription factor binding site position, and datasets like that presented here will begin to provide data that can help us uncover how this parameter dictates gene expression. Indeed, with binding site positions and energy matrices in hand, we can systematically move these binding sites and explore the implications for the level of gene expression, providing a systematic tool to understand the role of  binding-site position.  \\

\begin{table}[h]
\rowcolors{2}{lightgray}{white}
\begin{tabular}{c l l r}
\toprule
\rowcolor{darkgray}
Architecture & \shortstack{Total number \\ of promoters}          & \shortstack{Number of promoters  \\ with at least one newly \\ discovered binding site}  \\
\midrule
\hline
{All Architectures} & 113 & 52\\
{(0,0)} & 32 & 0\\
{(0,1)} & 26 & 23\\
{(1,0)} & 13 & 10\\
{(1,1)} & 6 & 6\\
{(0,2)} & 4 & 3\\
{(2,0)} & 6 & 5\\
{(2,1)} & 2 & 2\\
{(1,2)} & 1 & 1\\
{(2,2)} & 1 & 1\\
{(3,0)} & 3 & 1\\
{(0,4)} & 1 & 0\\
{inactive} & 18 & 0 \\
\bottomrule
\end{tabular}
\caption{\label{tab:allprom} All promoters examined in this study, categorized according to type of regulatory architecture. Those promoters which have no recognizable RNAP site are labeled as inactive rather than constitutively expressed (0, 0). }
\end{table}

Figure~\ref{fig:architectures} delves more deeply into the various regulatory architectures described in Figure \ref{fig:counts_summary}(B) by showing several example promoters for each of
the different architecture types. In each of the cases shown in the figure, prior to the work presented here, these promoters had no regulatory information in relevant databases such as  Ecocyc \cite{Keseler2016} and RegulonDB \cite{Santos_Zavaleta2019}.  Now, using the sequencing methods explained above  we were able to identify candidate binding sites.  For a number of cases, these putative binding sites were then used to synthesize oligonuceotide probes to enrich and identify their corresponding putative transcription factor using mass spectrometry. While Figure~\ref{fig:architectures} gives a sense of the kinds of regulatory architectures we discovered in this study, our entire collection of regulatory cartoons can be found in Supplementary Figure S6. \\

A recent paper christened that part of the {\it E. coli} genome for which the function of the genes is unknown the y-ome \cite{Ghatak2019}. Their surprising finding is that roughly 35\% of the genes in the  {\it E. coli} genome are functionally unannotated. The situation is likely worse for other organisms. For many of the genes in the y-ome, we remain similarly ignorant of how those genes are regulated. Figures \ref{fig:architectures} and \ref{fig:one_gene} provide several examples from the y-ome, of genes and transcription factors for which little to nothing was previously known. As shown in Figure \ref{fig:one_gene}, our study has found the first examples that we are aware of in the entire {\it E. coli} genome of a binding site for YciT. These examples are intended to show the outcome of the methods developed here and to serve as an invitation to browse the online resource (\url{www.rpgroup.caltech.edu/RNAseq_SortSeq/interactive_a})  to see many examples of the regulation of y-ome genes. \\  

\clearpage

\begin{table}[!h]
\rowcolors{2}{lightgray}{white}
\begin{tabular}{c l l l l l r}
\toprule
\rowcolor{darkgray}
Architecture & Promoter         & \shortstack{Newly \\ discovered \\ binding \\ sites} & \shortstack{Literature \\ binding \\ sites} & \shortstack{Identified \\ binding \\ sites} & Evidence\\
\midrule
\hline
{(0, 0)} & \it{acuI} & 0 & 0 & 0 &\\
& \it{adiY} & 0 & 0 & 0 &\\
& \it{arcB} & 0 & 0 & 0 &\\
& \it{coaA} & 0 & 0 & 0 &\\
& \it{dnaE} & 0 & 0 & 0 &\\
& \it{ecnB} & 0 & 0 & 0 &\\
& \it{holC} & 0 & 0 & 0 &\\
& \it{hslU} & 0 & 0 & 0 &\\
& \it{htrB} & 0 & 0 & 0 &\\
& \it{modE} & 0 & 0 & 0 &\\
& \it{motAB-cheAW} & 0 & 0 & 0 &\\
& \it{poxB} & 0 & 0 & 0 &\\
& \it{rcsF} & 0 & 0 & 0 &\\
& \it{rumB} & 0 & 0 & 0 &\\
& \it{sbcB} & 0 & 0 & 0 &\\
& \it{sdaB} & 0 & 0 & 0 &\\
& \it{ybdG} & 0 & 0 & 0 &\\
& \it{ybiP} & 0 & 0 & 0 &\\
& \it{ybjL} & 0 & 0 & 0 &\\
& \it{ybjT} & 0 & 0 & 0 &\\
& \it{yehS} & 0 & 0 & 0 &\\
& \it{yehT} & 0 & 0 & 0 &\\
& \it{yfhG} & 0 & 0 & 0 &\\
& \it{ygdH} & 0 & 0 & 0 &\\
& \it{ygeR} & 0 & 0 & 0 &\\
& \it{yggW} & 0 & 0 & 0 &\\
& \it{ygjP} & 0 & 0 & 0 &\\
& \it{ynaI} & 0 & 0 & 0 &\\
& \it{yqhC} & 0 & 0 & 0 &\\
& \it{zapB} & 0 & 0 & 0 &\\
& \it{zupT} & 0 & 0 & 0 &\\
& \it{amiC} & 0 & 0 & 0 &\\
{(0, 1)} & \it{aegA} & 1 & 0 & 0 &\\
& \it{bdcR} & 1 & 0 & 1 & \shortstack{Known binding \\ location (NsrR) \\ ~\cite{partridge_nsrr_2009-1}}\\
& \it{dicC} & 0 & 1 & 0 &\\
& \it{fdoH} & 1 & 0 & 0 &\\
& \it{groSL} & 1 & 0 & 0 &\\
& \it{idnK} & 1 & 0 & 1 & \shortstack{Mass- \\ Spectrometry (YgbI)}\\
& \it{leuABCD} & 1 & 0 & 1 & \shortstack{Mass- \\ Spectrometry (YgbI)}\\
\bottomrule
\end{tabular}
\caption{continued on next page}
\end{table}

\setcounter{table}{1}    

\begin{table}[!h]
\rowcolors{2}{lightgray}{white}
\begin{tabular}{c l l l l l r}
\toprule
\rowcolor{darkgray}
Architecture & Promoter          & \shortstack{Newly \\ discovered \\ binding \\ sites} & \shortstack{Literature \\ binding \\ sites} & \shortstack{Identified \\ binding \\ sites} & Evidence\\
\midrule
& \it{pcm} & 1 & 0 & 0 &\\
& \it{yedK} & 1 & 0 & 1 & \shortstack{Mass- \\ Spectrometry (TreR)}\\
& \it{rapA} & 1 & 0 & 1 & \shortstack{Growth condition \\ Knockout (GlpR),\\ Bioinformatic (GlpR)}\\
& \it{sdiA} & 1 & 0 & 0 &\\
& \it{tar} & 1 & 0 & 0 &\\
& \it{tff-rpsB-tsf}
 & 1 & 0 & 1 & \shortstack{Growth condition \\ Knockout (GlpR), \\ Bioinformatic (GlpR), \\ Knockout   (GlpR)}\\
& \it{thiM} & 1 & 0 & 0 &\\
& \it{tig} & 1 & 0 & 1 & \shortstack{Growth condition \\ Knockout (GlpR), \\ Bioinformatic (GlpR), \\ Knockout (GlpR)}\\
& \it{ycgB} & 1 & 0 & 0 &\\
& \it{ydjA} & 1 & 0 & 0 &\\
& \it{yedJ} & 1 & 0 & 0 &\\
& \it{ycbZ} & 1 & 0 & 0 &\\
& \it{phnA} & 1 & 0 & 1 & \shortstack{Mass- \\ Spectrometry (YciT)}\\
& \it{mutM} & 1 & 0 & 0 &\\
& \it{rhlE} & 1 & 0 & 1 & \shortstack{Growth condition \\ Knockout (GlpR), \\ Bioinformatic (GlpR), \\ Mass-\\
Spectrometry (GlpR)}\\
& \it{uvrD} & 1 & 0 & 1 & Bioinformatic (LexA)\\
& \it{dusC} & 1 & 0 & 0 &\\
& \it{ftsK} & 0 & 1 & 0 &\\
& \it{znuA} & 0 & 1 & 0 &\\
{(1, 0)} & \it{waaA-coaD} & 1 & 0 & 0 &\\
& \it{cra} & 1 & 0 & 0 & \\
& \it{iap} & 1 & 0 & 0 &\\
& \it{araC} & 0 & 1 & 0 & \\
& \it{minC} & 1 & 0 & 0 &\\
& \it{ybeZ} & 1 & 0 & 0 &\\
& \it{mscM} & 1 & 0 & 0 & \\
& \it{mscS} & 1 & 0 & 0 & \\
& \it{rlmA} & 1 & 0 & 0 & \\
& \it{thrLABC} & 1 & 0 & 0 & \\
& \it{yeiQ} & 1 & 0 & 1 & \shortstack{Growth condition \\ Knockout (FNR), \\ Bioinformatic (FNR)}\\
\bottomrule
\end{tabular}
\caption{continued on next page}
\end{table}

\setcounter{table}{1}   

\begin{table}[!h]
\rowcolors{2}{lightgray}{white}
\begin{tabular}{c l l l l l r}
\toprule
\rowcolor{darkgray}
Architecture & Promoter          & \shortstack{Newly \\ discovered \\ binding \\ sites} & \shortstack{Literature \\ binding \\ sites} & \shortstack{Identified \\ binding \\ sites} & Evidence\\
\midrule
& \it{dgoR} & 0 & 1 & 0 & \shortstack{Mass- \\ Spectrometry (DgoR)}\\
& \it{lac} & 0 & 1 & 0 & \shortstack{Mass- \\ Spectrometry (LacI)}\\
{(0, 2)} & \it{yecE} & 2 & 0 & 1 & \shortstack{Mass- \\ Spectrometry (HU)}\\
& \it{yjjJ} & 2 & 0 & 1 & \shortstack{Growth condition \\ Knockout (MarA), \\ Bioinformatic (MarA)}\\
& \it{dcm} & 2 & 0 & 1 & \shortstack{Mass- \\ Spectrometry (HNS)}\\
& \it{marR} & 0 & 2 & 0 & \shortstack{Mass- \\ Spectrometry (MarR)}\\
{(1, 1)} & \it{ilvC} & 2 & 0 & 1 & \shortstack{Mass- \\ Spectrometry (IlvY) \\ ~\cite{rhee_activation_1998} }\\
& \it{ybiO} & 2 & 0 & 0 & \\
& \it{yehU} & 2 & 0 & 1 & \shortstack{Growth condition \\ Knockout (FNR), \\ Bioinformatic (FNR)}\\
& \it{ykgE} & 2 & 0 & 2 & \shortstack{Growth condition \\ Knockout (FNR), \\ Bioinformatic (FNR), \\ Mass-\\
Spectrometry(YieP) \\ Knockout (YieP)}\\
& \it{ymgG} & 2 & 0 & 0 & \\
& \it{znuCB} & 1 & 1 & 0 & \\
{(2, 0)} & \it{aphA} & 2 & 0 & 2 & \shortstack{Growth condition \\ Knockout (FNR), \\ Bioinformatic (FNR), \\ Mass-\\
Spectrometry (DeoR)}\\
& \it{arcA} & 2 & 0 & 2 & \shortstack{Growth condition \\ Knockout (FNR), \\ Bioinformatic (FNR), \\ Mass-\\
Spectrometry (FNR, CpxR)}\\
& \it{asnA} & 2 & 0 & 0 &\\
& \it{fdhE} & 2 & 0 & 2 & \shortstack{Growth condition \\ Knockout (FNR, ArcA), \\ Bioinformatic (FNR, ArcA), \\ Knockout (ArcA)}\\
& \it{xylF} & 0 & 2 & 0 & \\
& \it{mscL} & 2 & 0 & 0 & \\
\bottomrule
\end{tabular}
\caption{continued on next page}
\end{table}

\setcounter{table}{1}   

\begin{table}[t]
\rowcolors{2}{lightgray}{white}
\begin{tabular}{c l l l l l r}
\toprule
\rowcolor{darkgray}
Architecture & Promoter          & \shortstack{Newly \\ discovered \\ binding \\ sites} & \shortstack{Literature \\ binding \\ sites} & \shortstack{Identified \\ binding \\ sites} & Evidence\\
\midrule
{(2, 1)}  & \it{maoP} & 3 & 0 & 3 & \shortstack{Growth condition \\ Knockout (GlpR), \\ Bioinformatic (GlpR),\\Knockout (PhoP, HdfR, GlpR)}\\
& \it{rspA} & 1 & 2 & 1 & \shortstack{Mass-\\
Spectrometry (DeoR)}\\
{(1, 2)} & \it{dinJ} & 3 & 0 & 0 & \\
{(2, 2)} & \it{ybjX} & 4 & 0 & 4 & \shortstack{Bioinformatic (PhoP),\\
Mass-\\
Spectrometry (HNS, StpA)}\\
{(3, 0)} & \it{araAB} & 0 & 3 & 0 & \\
& \it{xylA} & 0 & 3 & 0 & \\
& \it{yicI} & 3 & 0 & 0 & \\
{(0, 4)} & \it{relBE} & 0 & 4 & 0 & \shortstack{Mass- \\ Spectrometry (RelBE)}\\
\bottomrule
\end{tabular}
\caption{\label{tab:allgenes} All genes investigated in this study categorized according to their   regulatory architecture, given as (number of activators, number of repressors). The table also lists the number of newly discovered binding sites, previously known binding sites, and number of identified transcription factors. The evidence used  for the transcription factor identification is given in the final column.}
\end{table}

\clearpage

\begin{figure}[!h]
\centering
\includegraphics[width=5.0truein]{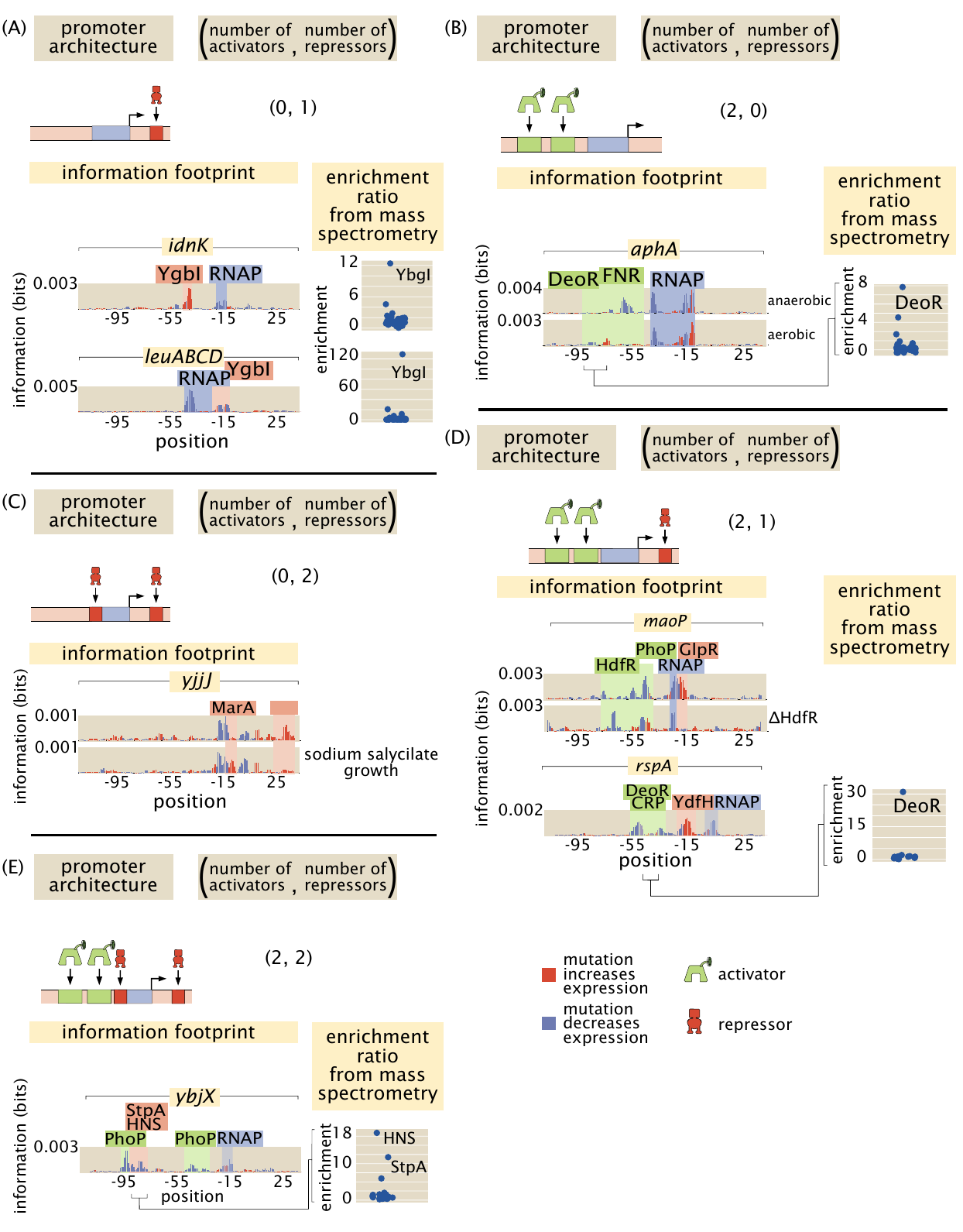}
\caption{Newly discovered or updated regulatory architectures. Examples of information footprints, gene knockouts, and mass spectrometry data used to identify transcription factors for five genes. (A) Examples of simple repression, i.e. (0, 1) architectures where the locations of the putative binding sites are highlighted in red and the identities of the bound transcription factors are revealed in the mass spectrometry data. (B) An example of a (2, 0) architecture. During aerobic growth FNR is inactive, but the DeoR site now has a significant effect on expression. (C) An example of a (0, 2) architecture. \textit{yjjJ} is regulated by MarA, which is only active in growth with sodium salycilate, and an unknown repressor. (D) An example of a (2, 1) architecture. (E) An example of a (2, 2) architecture.}
\label{fig:architectures}
\end{figure}

\clearpage

The ability to find binding sites for both widely acting regulators and transcription factors which may have only a few sites in the whole genome allows us to get an in-depth and quantitative view of any given promoter. As indicated in Figures~\ref{fig:one_gene}(A) and (B),  we were able to perform the relevant search and capture for the transcription factors that bind our putative binding sites.  In both of these cases, we now hypothesize that these newly discovered binding site-transcription factor pairs exert their control through repression. The ability to extract the quantitative features of regulatory control through energy matrices means that we can take a nearly unstudied gene such as \textit{ykgE}, which is regulated by an understudied transcription factor YieP, and quickly get to the point at which we can do quantitative modeling in the style that we and many others have performed  on the \textit{lac} operon ~\cite{Vilar2003a,Vilar2003b,bintu_transcriptional_2005,Kinney2010,garcia_quantitative_2011,Saiz2013,Barnes2019,Phillips2019}. \\

    One of the revealing case studies that demonstrates the broad reach of our approach for discovering regulatory architectures is offered by the insights we have gained into two widely acting regulators, GlpR ~\cite{schweizer_repressor_1985} and FNR ~\cite{korner_phylogeny_2003,kargeti_effect_2017}. In both cases, we have expanded the array of promoters that they are  now known to regulate. Further, these two case studies illustrate  that even for widely acting transcription factors, there is a large gap in regulatory knowledge and the approach advanced here has the power to discover new regulatory motifs. The newly discovered binding sites in Figure~\ref{fig:GlpR}(A) more than double the number of operons known to be regulated by GlpR as reported in RegulonDB \cite{Santos_Zavaleta2019}. We found 5 newly regulated operons in our data set, even though we were not specifically targeting GlpR regulation. Although the number of example promoters across the genome that we considered is too small to make good estimates, finding 5 regulated operons out of approximately 100 examined operons supports the claim that GlpR widely regulates and many more of its sites would be found in a full search of the genome. The regulatory roles revealed in Figure~\ref{fig:GlpR}(A) also reinforce the evidence that GlpR is a repressor. \\

For the GlpR-regulated operons newly discovered here, we found that this repressor binds strongly in the presence of glucose while all other growth conditions result in greatly diminished, but not entirely abolished, binding (Figure \ref{fig:GlpR}(A)). As there is no previously known direct molecular interaction between GlpR and glucose and the repression is reduced but not eliminated, the derepression in the absence of glucose is likely an indirect effect. As a potential mechanism of the indirect effect, \textit{gpsA} is known to be activated by CRP \cite{seoh_catabolic_1999}, and GpsA is involved in the synthesis of glycerol-3-phosphate (G3P), a known binding partner of GlpR which disables its repressive activity \cite{Larsons1987}. Thus in the presence of glucose GpsA and consequently G3P will be found in low concentration, ultimately allowing GlpR fulfill its role as a repressor. \\

Prior to this study, there were 4 operons known to be regulated by GlpR, each with between 4 and 8 GlpR binding sites \cite{Gama-Castro2016}, where the absence of glucose and the partial induction of GlpR was not enough to prompt a notable change in gene expression ~\cite{lin_glycerol}. These previously explored operons seemingly are regulated as part of an AND gate, where high G3P concentration \textit{and} an absence of glucose is required for high gene expression. By way of contrast, we have discovered operons whose regulation appears to be mediated by a single GlpR site per operon. With only a single site, GlpR functions as an indirect glucose sensor, as only the absence of glucose is needed to relieve repression by GlpR. \\

\begin{figure}[!h]
\centering
\includegraphics[width=4.5truein]{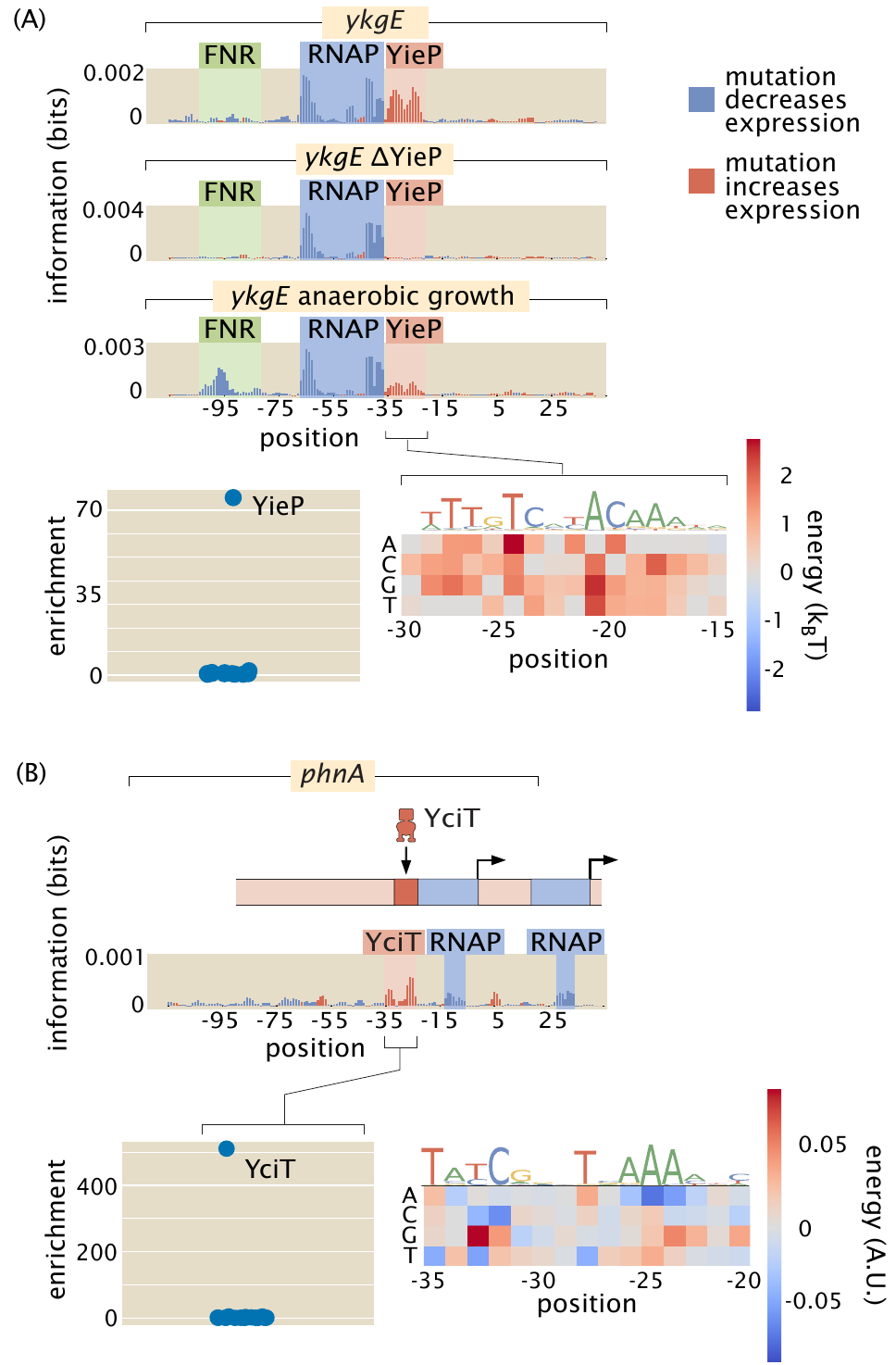}
\caption{Examples of the insight gained by Reg-Seq in the context of promoters with no previously
known regulatory information. (A) From the information footprint of the \textit{ykgE} promoter under different growth conditions, we can identify a repressor binding site downstream of the RNAP binding site. From the enrichment of proteins bound to the DNA sequence of the putative repressor as compared to a control sequence, we can identify YieP as the transcription factor bound to this site as it has a much higher enrichment ratio than any other protein. Lastly, the binding energy matrix for the repressor site along with corresponding sequence logo shows that the wild type sequence is the strongest possible binder and it displays an imperfect inverted repeat symmetry. (B) Illustration of  a comparable  dissection for the \textit{phnA} promoter.}
\label{fig:one_gene}
\end{figure}

\clearpage

\begin{figure}[!h]
\centering
\includegraphics[width=5.0truein]{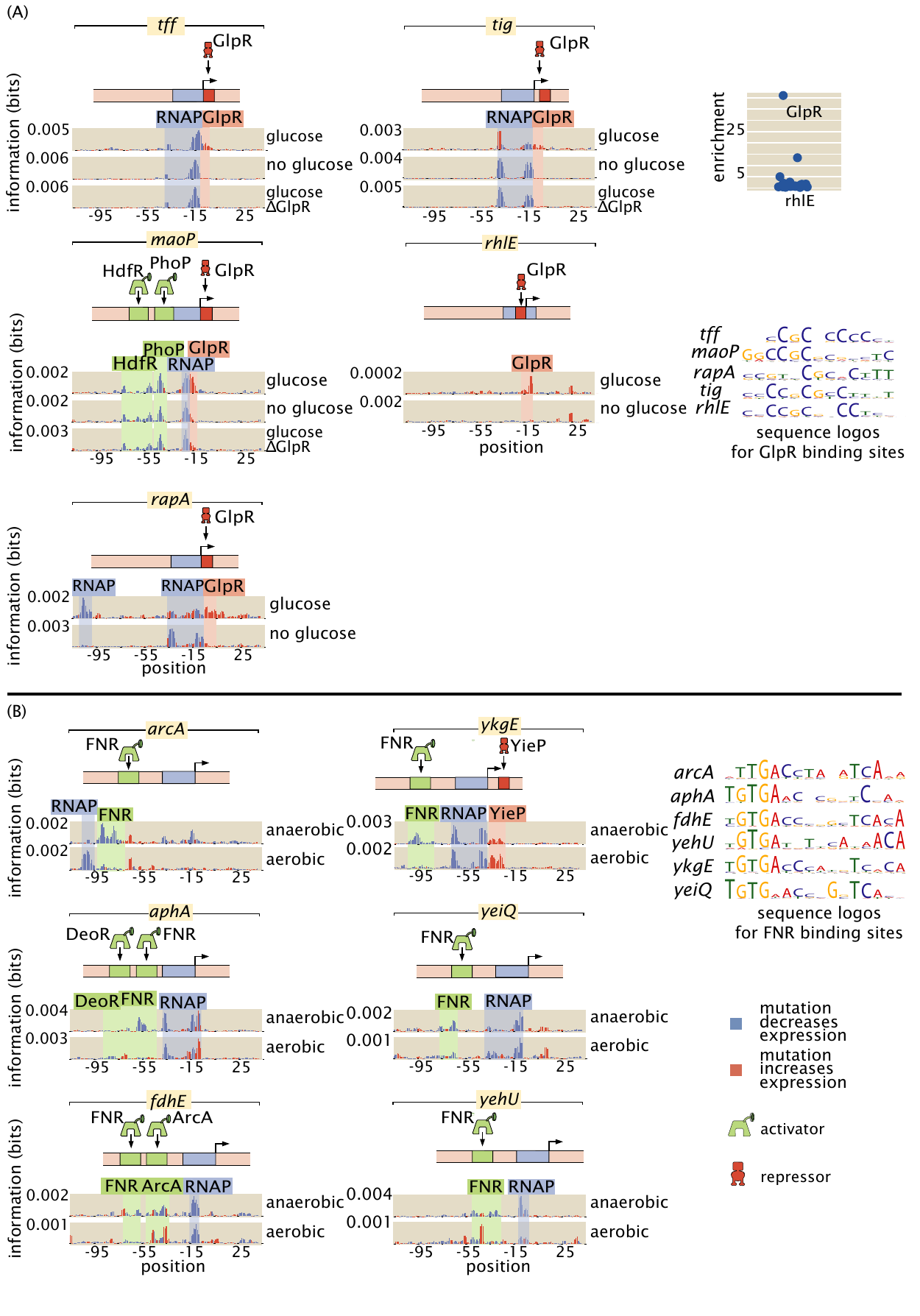}
\caption{Reg-Seq analysis of broadly-acting transcription factors.  (A) GlpR as a widely-acting regulator. Here we show the many promoters which we found to be regulated by GlpR, all of which were previously unknown. GlpR was demonstrated to bind to \textit{rhlE} by mass spectrometry enrichment experiments as shown in the top right. Binding sites in the \textit{tff}, \textit{tig}, \textit{maoP}, \textit{rhlE}, and \textit{rapA} have similar DNA binding preferences as seen in the sequence logos and each TF binding site binds strongly only in the presence of glucose. These similarities suggest that the same TF binds to each site. To test this hypothesis we knocked out GlpR and ran the Reg-Seq experiments for \textit{tff}, \textit{tig}, and \textit{maoP}.  We see that knocking out GlpR removes the binding signature of the TF. (B) FNR as a global regulator. FNR is known to be upregulated in anaerobic growth, and here we found it to regulate a suite of six genes. In growth conditions with prevalent oxygen the putative FNR sites are weakened, and the DNA binding preference of the six sites are shown to be similar from the sequence logos displayed on the right.}
\label{fig:GlpR}
\end{figure}

\clearpage

The second widely acting regulator our study revealed, FNR, has 151 binding sites already reported in RegulonDB and is well studied compared to most transcription factors ~\cite{Gama-Castro2016}. However, the newly discovered FNR sites displayed in Figure~\ref{fig:GlpR}(B) demonstrate that even for  well-understood transcription factors there is much still to be uncovered.  Our information footprints are in agreement with previous studies suggesting that FNR acts as an activator.  In the presence of O$_2$, dimeric FNR is converted to a monomeric form and its ability to bind DNA is greatly reduced  \cite{Myers2013}. Only in low oxygen conditions did we observe a binding signature from FNR, and we show a representative example of the information footprint from one of 11 growth conditions with plentiful oxygen in Figure~\ref{fig:GlpR}(B). \\

\begin{figure}[!h]
\centering
\includegraphics[width=6.5truein]{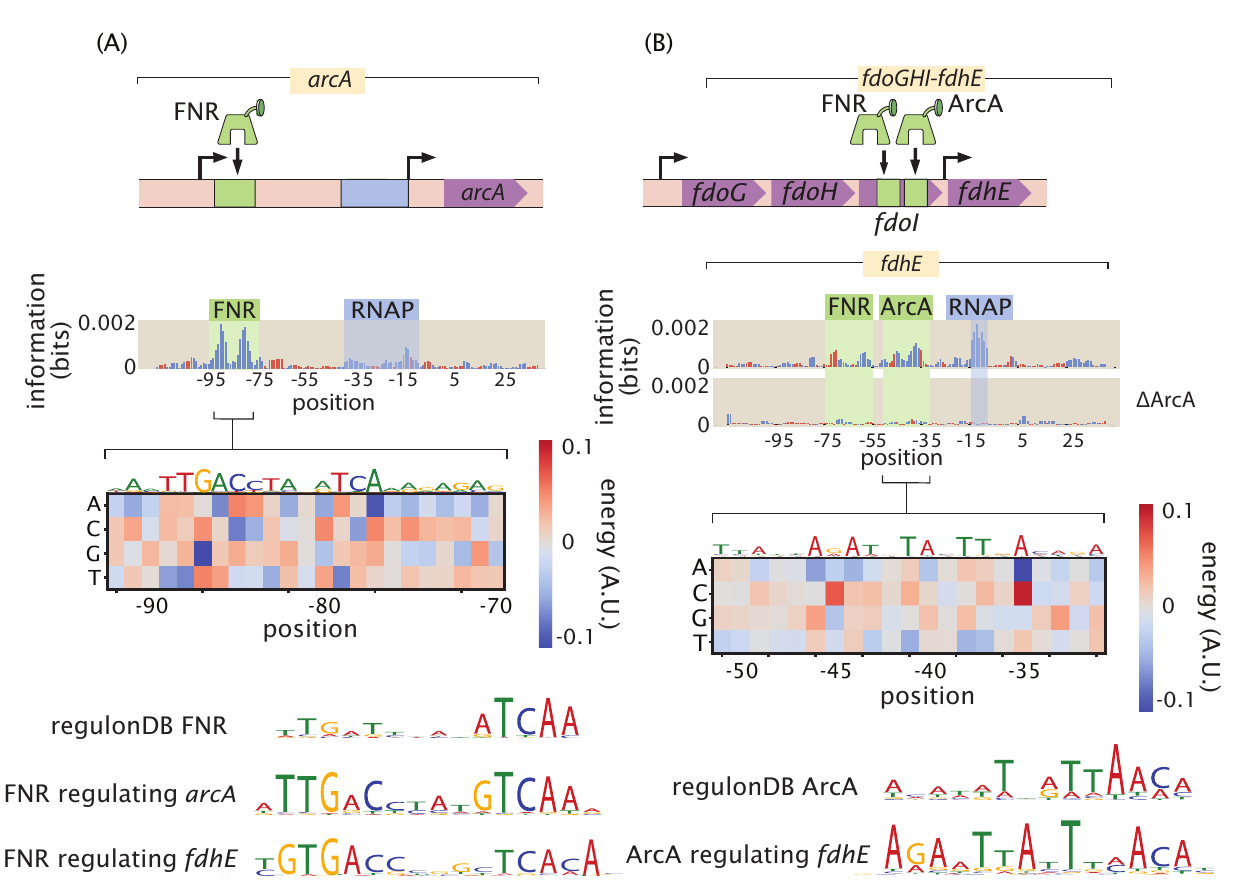}
\caption{Inspection of an anaerobic respiration genetic circuit. (A) Here we see not only how the \textit{arcA} promoter is regulated, but also the role this transcription factor plays in the regulation of another promoter. (B) Intra-operon regulation of \textit{fdhE} by both FNR and ArcA. A TOMTOM \cite{Gupta2007} search of the binding motif found that ArcA was the most likely candidate for the transcription factor. A knockout of ArcA demonstrates that the binding signature of the site, and its associated RNAP site, are no longer significant determinants of gene expression.}
\label{fig:fdhE}
\end{figure}

We observe quantitatively how FNR affects the expression of \textit{fdhE} both directly through transcription factor binding (Figure ~\ref{fig:fdhE}(A)) and indirectly through increased expression of ArcA (Figure \ref{fig:fdhE}(B)). Also, fully understanding even a single operon often requires investigating several regulatory regions as we have in the case of ~\textit{fdoGHI-fdhE} by investigating the main promoter for the operon as well as the promoter upstream of ~\textit{fdhE}. 36$\%$ of all multi-gene operons have at least one TSS which transcribes only a subset of the genes in the operon \cite{conway2014}.  Regulation within an operon is even more poorly studied than regulation in general. The main promoter for \textit{fdoGHI-fdhE} has a repressor binding site, which demonstrates that there is regulatory control of the entire operon. However, we also see in Figure \ref{fig:fdhE}(B) that there is control at the promoter level, as \textit{fdhE} is regulated by both ArcA and FNR and will therefore be upregulated in anaerobic conditions \cite{compan_anaerobic_1994}. The main TSS transcribes all four genes in the operon, while the secondary site shown in Figure \ref{fig:fdhE}(B) only transcribes \textit{fdhE}, and therefore anaerobic conditions will change the stoichiometry of the proteins produced by the operon. At the higher throughput that we use in this experiment it becomes feasible to target multiple promoters within an operon as we have done with \textit{fdoGHI-fdhE}. We can then determine under what conditions an operon is
internally regulated.  Figure~\ref{fig:fdhE} also makes it clear that for cases such as 
\textit{fdoGHI-fdhE}, there are many subtleties both in the interpretation of the information
footprints and in the construction of regulatory cartoons that are simultaneously accurate
and transparent.  A crucial next step in the development of these analyses is to move
from manual curation of the data to automated statistical analyses that can help make sense of
these complicated datasets. \\

\begin{figure}[!h]
\centering
\includegraphics[width=6.5truein]{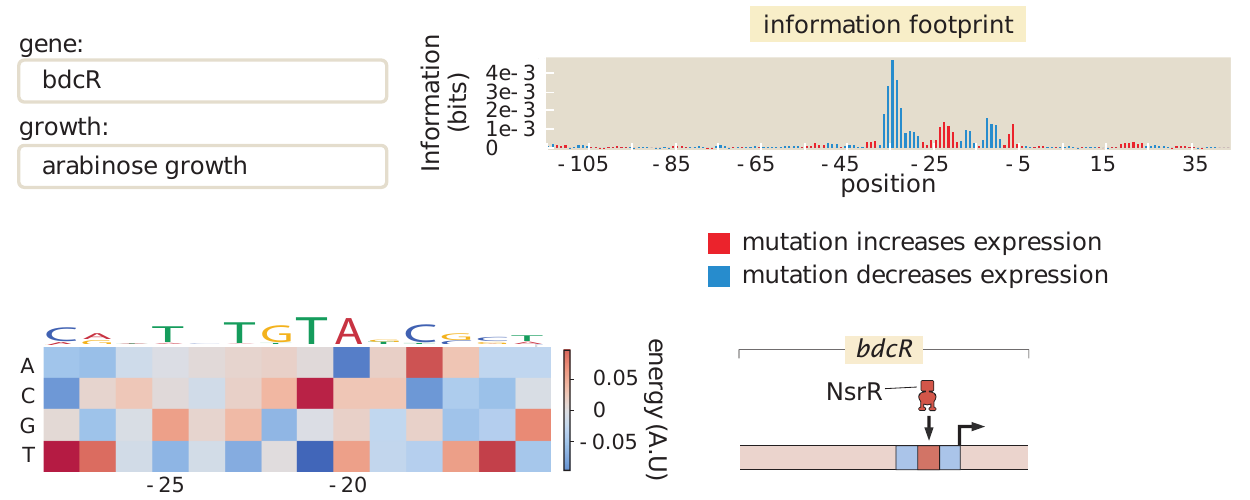}
\caption{Representative view of the interactive figure that is available online. This interactive figure  captures the entirety of our dataset. Each figure features a drop-down menu of genes and growth conditions. For each such gene and growth condition, there is a corresponding information footprint revealing putative binding sites, an energy matrix that shows the strength of binding of the relevant transcription factor to those binding sites and a cartoon that schematizes the newly-discovered regulatory architecture of that gene.}
\label{fig:interactive}
\end{figure}

By examining the over 100 promoters considered here, grown under 12 growth conditions, we have a total of more than 1000 information footprints and data sets. In this age of big data, methods to explore and draw insights from that data are crucial.  To that end, as introduced in Figure~\ref{fig:interactive}, we have developed an online resource (see \url{www.rpgroup.caltech.edu/RNAseq_SortSeq/interactive_a}) that makes it possible for anyone who is interested to view our data and draw their own biological conclusions. Information footprints for any combination of gene and growth condition are displayed via drop down menus. Each identified transcription factor binding site or transcription start site is marked, and energy matrices for all transcription factor binding sites are displayed. In addition, for each gene, we feature a simple cartoon-level schematic that captures our now current best understanding of the regulatory architecture and resulting mechanism. \\

The interactive figure in question was invaluable in identifying transcription factors, such as GlpR, whose binding properties vary depending on growth condition. As sigma factor availability also varies greatly depending on growth condition, studying the interactive figure identified many of the secondary RNAP sites present. The interactive figure provides a valuable resource both to those who are interested in the regulation of a particular gene and those who wish to look for patterns in gene regulation across multiple genes or across different growth conditions. \\

\section{Discussion} \label{discussion} 

The study of gene regulation is one of the centerpieces of modern biology. As a result, it is surprising that in the genome era, our ignorance of the regulatory landscape in even the best-understood model organisms remains so vast. Despite understanding the regulation of transcription initiation in bacterial promoters \cite{Browning2016}, and how to tune their expression, we lack an experimental framework to unravel understudied promoter architectures at scale.  As such, in our view one of the grand challenges of the genome era is the need to uncover the regulatory landscape for each and every organism with a known genome sequence. Given the ability to read and write DNA sequence at will, we are convinced that to make that reading of DNA sequence  truly informative about biological function and to give that writing the full power and poetry of what Crick christened ``the two great polymer languages'', we need a full accounting of how the genes of a given organism are regulated and how environmental signals communicate with the transcription factors that mediate that regulation -- the so-called ``allosterome'' problem \cite{Lindsley2006}. The work presented here provides a general methodology for making progress on the former problem and also demonstrates that,  by performing Reg-Seq in different growth conditions, we can make headway on the latter problem as well. \\

The advent of cheap DNA sequencing offers the promise of beginning to achieve that grand challenge goal in the form of MPRAs reviewed in \cite{Kinney2019}. A particular implementation of such methods was christened Sort-Seq \cite{Kinney2010} and was demonstrated in the context of well understood regulatory architectures. A second generation of the Sort-Seq method \cite{Belliveau2018} established experiments through the use of DNA-affinity chromatography and mass spectrometry which made it possible to identify the transcription factors that bind the putative binding sites discovered by Sort-Seq. But there were critical shortcomings in the method, not least of which was that it lacked the scalability to uncover the regulatory genome on a genome-wide basis. \\

The work presented here builds on the foundations laid in the previous studies by invoking RNA-Seq as a readout of the level of expression of the promoter mutant libraries needed to infer information footprints and their corresponding energy matrices and sequence logos followed by a combination of mass spectrometry and gene knockouts to identify the transcription factors that bind those sites. The case studies described in the main text showcase the ability of the method to deliver on the promise of beginning to uncover the regulatory genome systematically. The extensive online resources hint at a way of systematically reporting those insights in a way that can be used by the community at large to develop regulatory intuition for biological function and to design novel regulatory architectures using energy matrices. \\

However, several shortcomings remain in the approach introduced here. First, the current implementation of Reg-Seq still largely relies on manual curation as the basis of using information
footprints to generate testable regulatory hypotheses.  As described in the methods section, we have also used statistical testing as a way to convert information footprints into regulatory hypotheses, 
but there clearly remains  much work to be done on the data analysis pipeline to improve both the power and the accuracy of this approach.
In addition, these regulatory hypotheses can also be converted into gene regulatory models using statistical physics~\cite{Buchler2003,bintu_transcriptional_2005}. However, here too, 
as the complexity of the regulatory architectures increases, it will be of great interest to use automated model generation as suggested in a recent biophysically-based neural network approach~\cite{Tareen2019}.\\

A second key challenge faced by the methods described here is that the mass spectrometry and the gene knockout confirmation aspects of the experimental pipeline remain low-throughput. To overcome this, we have begun to explore a new generation of experiments such as {\it in vitro} binding assays  that will make it possible to accomplish transcription factor identification at higher throughput. Specifically, we are exploring multiplexed mass spectrometry measurements and multiplexed Reg-Seq on libraries of gene knockouts as ways to break the identification bottleneck. \\

Another shortcoming of the current implementation of the method is that it would miss regulatory action at a distance. Indeed, our laboratory has invested  a significant effort in exploring such long-distance regulatory action in the form of DNA looping in bacteria and VDJ recombination in jawed vertebrates. It is well known that transcriptional control through enhancers in eukaryotic regulation is central in contexts ranging from embryonic development to hematopoiesis \cite{Melnikov2012}. The current incarnation of the methods described here have focused on contiguous regions in the vicinity of the transcription start site. Clearly, to go further in dissecting the entire regulatory genome, these methods will have to be extended to non-contiguous regions of the genome.\\ 

The findings from this study provide a foundation for systematically performing genome-wide regulatory dissections. We have developed a method to pass from complete regulatory ignorance to designable regulatory architectures  and we are hopeful that others will adopt these methods with the ambition of uncovering the regulatory architectures that preside over their organisms of interest. \\

\section{Methods} \label{methods} 

\subsection{Library construction}

   Promoter variants were synthesized on a microarray (TWIST Bioscience, San Francisco, CA). The sequences were designed computationally such that each base in the 160 bp promtoter region has a 10$\%$ probability of being mutated. For each given promoter's library, we ensured that the mutation rate as averaged across all sequences was kept between 9.5$\%$ and 10.5$\%$, otherwise the library was regenerated. There are an average of 2200 unique promoter sequences per gene (for an analysis of how our results depend upon number of unique promoter sequences see Supplementary Figure S3). An average of 5 unique 20 base pair barcodes per variant promoter was used for the purpose of counting transcripts. The barcode was inserted 110 base pairs from the 5' end of the mRNA, containing 45 base pairs from the targeted regulatory region, 64 base pairs containing primer sites used in the construction of the plasmid, and 11 base pairs containing a three frame stop codon. All the sequences are listed in Supplementary Table 1. Following the barcode there is an RBS and a GFP coding region. Mutated promoters were PCR amplified and inserted by Gibson assembly into the plasmid backbone of pJK14 (SC101 origin) ~\cite{Kinney2010}. Constructs were electroporated into \textit{E. coli} K-12 MG1655 ~\cite{blattner_complete_1997}.  \\

\subsection{RNA preparation and sequencing}

    Cells were grown to an optical density of 0.3 and RNA was then stabilized using Qiagen RNA Protect  (Qiagen, Hilden, Germany). Lysis was performed using lysozyme (Sigma Aldrich, Saint Louis, MO) and RNA was isolated using the Qiagen RNA Mini Kit. Reverse transcription was preformed using Superscript IV (Invitrogen, Carlsbad, CA) and a specific primer for the labeled mRNA. qPCR was preformed to check the level of DNA contamination and the mRNA tags were PCR amplified and Illumina sequenced. Within a single growth condition, all promoter variants for all regulatory regions were tested in a single multiplexed RNA-Seq experiment. All sequencing was carried out by either the Millard and Muriel Jacobs Genetics and Genomics Laboratory at Caltech (HiSeq 2500) on a 100 bp single read flow cell or using the sequencing services from NGX Bio on a 250 bp or 150 base paired end flow cell. \\

\subsection{Analysis of sequencing results}
    To determine putative transcription factor binding sites, we first compute the effect of mutations on gene expression at a base pair-by-base pair level using information footprints. The information footprints are a hypothesis generating tool and we choose which regions to further investigate using techniques such as mass spectrometry by visually inspecting the data for regions of 10 to 20 base pairs that have high information content compared to background. Our technique currently relies on using human intuition to determine binding sites, but to validate these choices and to capture all regions important for gene expression we computationally identify regions where gene expression is changed significantly up or down by mutation (p $<$ 0.01), and discard any potential sites which do not fit this criteria.  We infer the effect of mutation using Markov Chain Monte Carlo, and we use the distribution of parameters from the inference to form a 99 $\%$ confidence interval for the average effect of mutation across a 15 base pair region. We include binding sites that are statistically significant at the 0.01 level in any of the tested growth conditions. \\
    
    Many false positives will be secondary RNAP sites and we remove from consideration any sites that resemble RNAP sites. We fit energy matrices to each of the possible binding sites and use the preferred DNA sequence for binding to identify the RNAP sites. We use both visual inspection to compare the preferred sequence to known consensus sequences for each of the \textit{E. coli} sigma factor binding sites (for example, do the preferred bases in the energy matrix have few mismatches to the TGNTATAAT extended minus 10 for $\sigma^{70}$ sites), and the TOMTOM tool \cite{Gupta2007} to computationally compare the potential site to examples of $\sigma^{70}$, $\sigma^{38}$, and $\sigma^{54}$ sites that we determined in this experiment. For further details see Supplementary Figure S4.  We discard any sites that have a p-value of similarity with an RNAP site of less than $5x10^{-3}$ in the TOMTOM analysis or are deemed to be too visually similar to RNAP sites. If a single site contains an RNAP site along with a transcription factor site we remove only those bases containing the probable RNAP site. This results in 95 identified transcription factor binding regions.\\
    
    For primary RNAP sites, we include a list of probable sigma factor identities as Supplementary Table 2. Sites are judged by visual similarity to consensus binding sites. Those sites where the true sigma factor is unclear due to overlapping binding sites are omitted. Overlapping binding sites  (from multiple TFs or RNAP sites) in general can pose issues for this method. In many cases, looking at growth conditions where only one of the relevant transcription factors is present or active is an effective way to establish site boundaries and infer correct energy matrices. For sites where no adequate growth condition can be found, or when a TF overlaps with an RNAP site, the energy matrix will not be reflective of the true DNA-protein interaction energies. If the TFs in overlapping sites are composed of one activator and one repressor, then we use the point at which the effect of mutation shifts from activator-like to repressor-like as a demarcation point between binding sites. We see a case of a potentially overlooked repressor due to overlapping sites in Figure~\ref{fig:architectures}(B), as there are several repressor like bases overlapping the RNAP -10 site and the effect weakens in low oxygen growth. However, due to the effect of the RNAP site, when averaged over a potential 15 base pair region, the repressor-like bases do not have a significant effect on gene expression.  \\

\subsection{DNA affinity chromatography and mass spectrometry}

    Upon identifying a putative transcription factor binding site, we used DNA affinity chromatography, as done in \cite{Belliveau2018} to isolate and enrich for the transcription factor of interest. In brief, we order biotinylated oligos of our binding site of interest (Integrated DNA Technologies, Coralville, IA) along with a control, "scrambled" sequence, that we expect to have no specificity for the given transcription factor. We tether these oligos to magnetic streptavidin beads (Dynabeads MyOne T1; ThermoFisher, Waltham, MA), and incubate them overnight with whole cell lysate grown in the presences of either heavy (with $^{15}$N) or light (with $^{14}$N) lysine for the experimental and control sequences, respectively. The next day, proteins are recovered by digesting the DNA with the PtsI restriction enzyme (New England Biolabs, Ipswich, MA), whose cut site was incorporated into all designed oligos. \\

    Protein samples were then prepared for mass spectrometry by either in-gel or in-solution digestion using the Lys-C protease (Wako Chemicals, Osaka, Japan). Liquid chromatography coupled mass spectrometry (LC-MS) was performed as previously described by ~\cite{Belliveau2018}, and is further discussed in the SI. SILAC labeling was performed by growing cells ($\Delta$ LysA) in either heavy isotope form of lysine or its natural form. \\

    It is also important to note that while we relied on the SILAC method to identify the TF identity for each promoter, our approach doesn’t require this specific technique. Specifically, our method only requires a way to contrast between the copy number of proteins bound to a target promoter in relation to a scrambled version of the promoter. In principle, one could use multiplexed proteomics based on isobaric mass tags \cite{Pappireddi2019} to characterize up to 10 promoters in parallel. Isobaric tags are reagents used to covalently modify peptides by using the heavy-isotope distribution in the tag to encode different conditions. The most widely adopted methods for isobaric tagging are the isobaric tag for relative and absolute quantitation (iTRAQ) and the tandem mass tag (TMT). This multiplexed approach involves the fragmentation of peptide ions by colliding with an inert gas. The resulting ions are resolved in a second MS-MS scan (MS2). \\
    
    Only a subset (13) of all transcription factor targets were identified by mass spectrometry due to limitations in scaling the technique to large numbers of targets. The transcription factors identified by this method are enriched more than any other DNA binding protein, with p $<$ 0.01 using the outlier detection method as outlined by Cox and Mann~\cite{cox_maxquant_2008}, with corrections for multiple hypothesis testing using the method proposed by Benjamini and Hochberg~\cite{benjamini_controlling_1995}. \\

\subsection{Construction of knockout strains}

    Conducting DNA affinity chromatography followed by mass spectrometry on putative binding sites resulted in potential candidates for the transcription factors that are responsible for the information contained at a given promoter region. For some cases, to verify that a given transcription factor is, in fact, regulating a given promoter, we repeated the RNA sequencing experiments on strains with the transcription factor of interest knocked out. \\

    To construct the knockout strains, we ordered strains from the Keio collection \cite{Yamamoto2009} from the Coli Genetic Stock Center. These knockouts were put in a MG1655 background via phage P1 transduction and verified with Sanger sequencing. To remove the kanamycin resistance that comes with the strains from the Keio collection, we transformed in the pCP20 plasmid, induced FLP recombinase, and then selected for colonies that no longer grew on either kanamycin or ampicillin. Finally, we transformed our desired promoter libraries into the constructed knockout strains, allowing us to perform the RNA sequencing in the same context as the original experiments. \\
    
\subsection{Code and Data Availability}
All code used for processing data and
plotting as well as the final processed data, plasmid sequences, and primer
sequences can be found on the GitHub repository (\url{https://github.com/RPGroup-PBoC/RNAseq_SortSeq}) doi:10.5281/zenodo.3611914.
 Energy matrices were generated using the MPAthic software ~\cite{ireland_mpathic_2016}.
All raw sequencing data is available at
the Sequence Read Archive (accession no.PRJNA599253). All inferred information footprints and energy matrices can be found on the CalTech data repository doi:10.22002/D1.1331. All mass spectrometry raw data is available on the CalTech data repository doi:10.22002/d1.1336

\section{Acknowledgments} 

We are grateful to Rachel Banks, Stephanie Barnes,  Curt Callan, Griffin Chure, Ana Duarte, Vahe Galstyan, Hernan Garcia, Soichi Hirokawa, Thomas Lecuit, Heun Jin Lee, Madhav Mani, Nicholas McCarty, Muir Morrison, Steve Quake, Tom R{\"o}schinger, Manuel Razo-Mejia,  Gabe Salmon, and Guillaume Urtecho for useful discussion and feedback on the manuscript. Guillaume Urtecho and Sri Kosuri have been instrumental in providing key advice and protocols at various stages in the development of this work.  We would like to thank Jost Vielmetter and Nina Budaeva for providing access to their Cell Disruptor. Brett Lomenick provided crucial help and advice with protein preparation. We also thank Igor Antoshechkin for his help with sequencing at the Caltech Genomics Facility.  \\ 

Funding: We are deeply grateful for support from NIH Grants DP1 OD000217 (Director's Pioneer Award) and  1R35 GM118043-01 (Maximizing Investigators Research Award) which made it possible to undertake this multi-year project. N.M.B. was supported by an HHMI International Student Research Fellowship. S.M.B was supported by the NIH Institutional National Research Service Award (5T32GM007616-38) provided through Caltech. The Proteome Exploration Laboratory is supported by, the Beckman Institute, and NIH 1S10OD02001301.\\
\bibliography{mainbib.bib}

\clearpage

\renewcommand{\figurename}{Supplementary Figure}
\setcounter{figure}{0}

\textbf{{\Large Supplementary Information for
``Deciphering the regulatory genome of \textit{Escherichia
coli}, one hundred promoters at a time''}}

\tableofcontents



\section{Extended details of experimental design} \label{SI:methods} 

\subsection{Choosing target genes} \label{genechoice}
    Genes in this study were chosen to cover several different categories. 29 genes had some information on their regulation already known to validate our method under a number of conditions. 37 were chosen because the work of \cite{Schmidt2015} demonstrated that gene expression changed significantly under different growth conditions. A handful of genes such as \textit{minC}, \textit{maoP}, or \textit{fdhE} were chosen because we found either their physiological significance interesting, as in the case of the cell division gene \textit{minC} or that we found the gene regulatory question interesting, such for the intra-operon regulation demonstrated by \textit{fdhE}. The remainder of the genes were chosen because they had no regulatory information, often had minimal information about the function of the gene, and had an annotated transcription start site (TSS) in RegulonDB. 

\subsection{Choosing transcription start sites} \label{TSS}
    A known limitation of the experiment is that the mutational window is limited to 160 bp. As such, it is important to correctly target the mutation window to the location around the most active TSS. To do this we first prioritized those TSS which have been extensively experimentally validated and catalogued in RegulonDB. Secondly we selected those sites which had evidence of active transcription from RACE experiments \cite{mendoza-vargas_genome-wide_2009} and were listed in RegulonDB. If the intergenic region was small enough, we covered the entire region with our mutation window. If none of these options were available, we used computationally predicted start sites. 

\subsection{Sequencing} \label{SeqAnalysis}
    All sequencing was carried out by either the Millard and Muriel Jacobs Genetics and Genomics Laboratory at Caltech (HiSeq 2500) on a 100 bp single read flow cell or using the sequencing services from NGX Bio on a 250 bp or 150 base paired end flow cell. The total library was first sequenced by PCR amplifying the region containing the variant promoters as well as the corresponding barcodes. This allowed us to uniquely associate each random 20 bp barcode with a promoter variant. Any barcode which was associated with a promoter variant with insertions or deletions was removed from further analysis. Similarly, any barcode that was associated with multiple promoter variants was also removed from the analysis. The paired end reads from this sequencing step were then assembled using the FLASH tool \cite{magoc_flash_2011}. Any sequence with PHRED score less than 20 was removed using the FastX toolkit. Additionally, when sequencing the initial library, sequences which only appear in the dataset once were not included in further analysis in order to remove possible sequencing errors. \\

    For all the MPRA experiments, only the region containing the random 20 bp barcode was sequenced, since the barcode can be matched to a specific promoter variant using the initial library sequencing run described above. For a given growth condition, each promoter yielded 50,000 to 500,000 usable sequencing reads. Under some growth conditions, genes were not analyzed further if they did not have at least 50,000 reads. \\ 
    
    To determine which base pair regions were statistically significant a 99$\%$ confidence interval was constructed using the MCMC inference to determine the uncertainty.
    
\subsection{Growth conditions}
     The growth conditions studied in this study were inspired
     by \cite{Schmidt2015} and  include differing carbon sources such as growth in M9 with $0.5 \%$ Glucose, M9 with acetate ($0.5 \%$), M9 with arabinose ($0.5\%$), M9 with Xylose ($0.5\%$) and arabinose ($0.5\%$), M9 with succinate ($0.5\%$), M9 with fumarate ($0.5\%$), M9 with Trehalose ($0.5\%$), and LB. In each case cell harvesting was done at an OD of 0.3. These growth conditions were chosen so as to span a wide range of growth rates, as well as to illuminate any carbon source specific regulators. \\
     
     We also used several stress conditions such as heat shock, where cells were grown in M9 and were subjected to a heat shock of 42 degrees for 5 minutes before harvesting RNA. We grew in low oxygen conditions. Cells were grown in LB in a container with minimal oxygen, although some will be present as no anaerobic chamber was used. This level of oxygen stress was still sufficient to activate FNR binding, and so activated the anaerobic metabolism. We also grew cells in M9 with Glucose and 5mM sodium salycilate. \\
     
     Growth with zinc was preformed at a concentration of 5mM ZnCl$_2$ and growth with iron was preformed by first growing cells to an OD of 0.3 and then adding FeCL$_2$ to a concentration of 5mM and harvesting RNA after 10 minutes. Growth without cAMP was accomplished by the use of the JK10 strain which does not maintain its cAMP levels. \\
     
     All knockout experiment were preformed in M9 with Glucose except for the knockouts for \textit{arcA}, \textit{hdfR}, and \textit{phoP} which were grown in LB. \\ 


\section{Validating Reg-Seq against previous methods and results}\label{SI:validation} 

The work presented here is effectively a third-generation of the use of Sort-Seq methods for the discovery of regulatory architecture.  The primary difference between the present work and previous generations \cite{Kinney2010, Belliveau2018} is the use of RNA-Seq rather than fluorescence and cell sorting as a readout of the level of expression of our promoter libraries. As such, there are many important questions to be asked about the comparison between the earlier methods and this work. We attack that question in several ways. First, as shown in Figure~\ref{fig:compfig}, we have performed a head-to-head comparison of the two approaches to be described further in this section. Second, as shown in the next section, our list of candidate promoters included roughly 20\% for which the community has some knowledge of their regulatory architecture. In these cases, we examined the extent to which our methods recover the known features of regulatory control about those promoters. \\

   \subsection{Comparison between Reg-Seq by RNA-Seq and fluorescent sorting}

    As the basis for comparing the results of the fluorescence-based Sort-Seq approach with our RNA-Seq-based approach, we use  information footprints, expression shifts and sequence logos as our metrics. Figure~\ref{fig:compfig} shows examples of this comparison for four distinct genes of interest. Figure~\ref{fig:compfig}(A) shows the results of the two methods for the \textit{lacZYA} promoter with special reference to the CRP binding site. Both the information footprint and the sequence logo identify the same binding site. \\

    Figure~\ref{fig:compfig}(B) provides a similar analysis for the \textit{dgoRKADT} promoter where once again the information footprints and the sequence logos from the two methods are in reasonable accord. Figure~\ref{fig:compfig}(C) provides a quantitative dissection of the \textit{relBE} promoter which is repressed by RelBE.  Here we use both information footprints and expression shifts as a way to quantify the significance of mutations to different binding sites across the promoter. Finally, Figure~\ref{fig:compfig}(D) shows a comparison of the two methods for the \textit{marRAB} promoter. The two approaches both identify a MarR binding site. \\

\begin{figure}
\centering
\includegraphics[width=6.5truein]{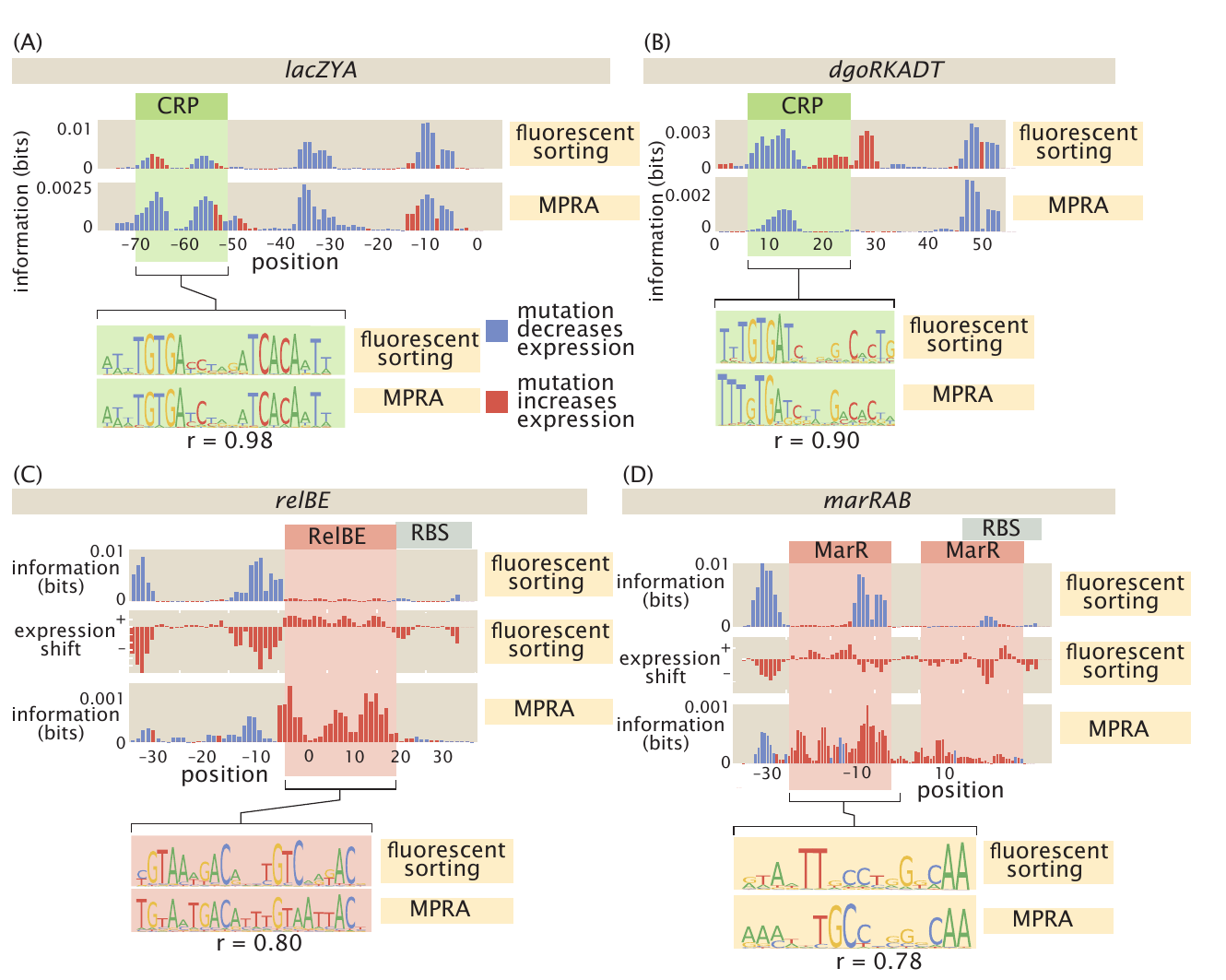}
\caption{A summary of four direct comparisons of measurements using fluorescence and sorting and using RNA-Seq. (A) CRP binds upstream of RNAP in the \textit{lacZYA} promoter. Despite the different measurement techniques for the two inferred energy matrices and their corresponding sequence logos, the CRP binding sites have a Pearson correlation coefficient of $r = 0.98$.  (B) The \textit{dgoRKADT} promoter is activated by CRP in the presence of galactonate. The FACS measurements were taken in the JK10 strain in the presence of 500mM cAMP. In both cases, a type II activator binding site can be identified based on the signals in the information footprint in the area indicated in green. Additionally the quantitative agreement between the CRP binding preference matrices are strong, with $r = 0.9$. (C) The \textit{relBE} promoter is repressed by RelBE. The inferred matrices between the two measurement methods have $r = 0.8$. (D) The \textit{marRAB} promoter is repressed by MarR. The features we can observe in the information footprint reflect this under measurement with both FACS or RNAseq. The inferred energy matrices (data not shown)  and sequence logos shown have $r = 0.78$. The right most MarR site overlaps with a ribosome binding site. The overlap has a stronger obscuring effect on the sequence specificity of the FACS measurement, which measures protein levels directly, than it does on the output of the RNAseq measurement.}
\label{fig:compfig}
\end{figure}

\subsection{Ability of Reg-Seq to recover known regulatory architectures}

In total, we have tested over 20 genes for which there is already some substantial regulatory knowledge reported in the literature. The successes and failures of this test are detailed in Figure \ref{fig:known_sites}. For those promoters which have strong evidence of a binding site, as determined by RegulonDB \cite{Santos_Zavaleta2019}, we recover all relevant transcription factor binding sites for 12 out of 16 cases, the majority of relevant binding sites for 2 out of 16 cases, and miss all or most of the regulation for just 2 promoters. We identify a total of 22 previously known high evidence binding sites. \\

\begin{figure}
\centering
   \includegraphics[width=6.5truein]{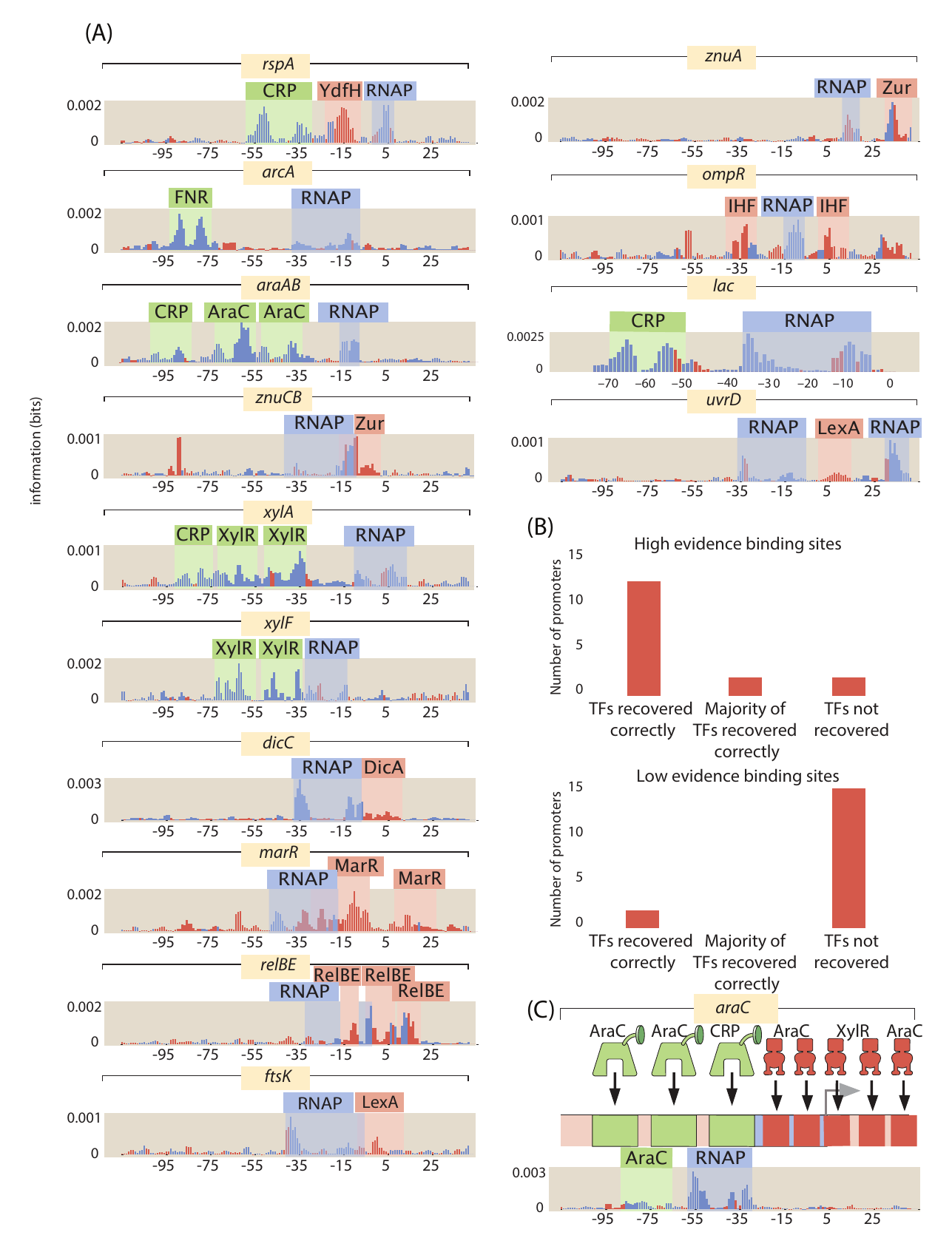}
    \caption{Reg-Seq analysis of ``gold standard'' promoters. (A) Information footprints for known and properly recovered binding sites. (B) A summary of how well the Reg-Seq results conform to literature results. The sites that are low evidence in the literature are determined by RegulonDB \cite{Santos_Zavaleta2019}. (C) The information footprint and known binding sites for the \textit{araC} promoter. Despite all the binding sites present, the only binding signature that appears is for RNAP.}
\label{fig:known_sites}
\end{figure}

    These results showcase that our method largely agrees with the established literature but also highlights several areas in which our method is prone to missing regulatory elements. One failure mode is caused by the presence of strong secondary binding sites. For example, in the \textit{araC} promoter, as shown in Figure \ref{fig:known_sites}(C), the only binding signatures that appear in the information footprint are from a secondary RNAP site. The secondary site seems to be expressed constitutively, and in the cases where the primary start site is even partially repressed, the secondary start site will dominate transcription and obscure the many binding sites that are in this promoter. \\
   
    If there are large numbers of regulatory elements, the data will often only show the few most important elements. If we look at the \textit{marR} promoter in Figure \ref{fig:known_sites}(C), we can only see the signature of the two MarR sites even though CpxR, Fis, and CRP are all known to bind to the promoter. MarR is a strong enough repressor that mutating any of the other transcription factor sites is unlikely to meaningfully change gene expression unless the MarR site is also mutated. This illustrates that the regulatory architectures discovered in this study represent a lower bound on what exists in each promoter. \\ 
   
   Finally, for some genes such as \textit{dicA} there was no known TSS prior to the experiment. Although there is a small regulatory region between \textit{dicA} and its neighboring gene, this does not ensure that we will include the strongest RNAP sites. Better mapping of transcription start sites could improve our method. \\ 
   
   We next consider low evidence binding sites. Other research determined the locations of the low evidence sites through gene expression analysis and sequence comparison to consensus sequences ~\cite{compan_anaerobic_1994,kumar_transcriptional_2011,easton_transcription_1983}. For 5 promoters in our list, the binding sites location itself is not known, only that the TF in question regulates the gene. For these promoters we recover the known regulation in only 2 out of 15 cases. Comparison to consensus sequences can be unreliable and generate false positives when the entirety of the \textit{E. coli} genome is considered. Gene expression analysis alone has difficulty ruling out indirect effects of a given transcription factor on gene expression and regulation determined by this method may occur outside of the 160 bp mutation window we consider. As our results recover high evidence sites well, the poor recovery of sites based on sequence gazing and gene expression analysis most likely indicates that these methods are unreliable for determining binding locations. \\

    We note that the first aim of our methods is regulatory discovery. We would like to be able to determine how previously uncharacterized promoters are regulated and ultimately, this is a question of binding-site and transcription factor identification. For that task, we do not require perfect correspondence between the two methods. With regulatory sites identified, our next objective is the determination of energy matrices that will allow us to turn binding site strength into a tunable knob that can nearly continuously tune the strength of transcription factor binding, thus altering gene expression in predictable ways as already shown in our earlier work \cite{Barnes2019}. The r-values between energy matrices range from 0.78 to 0.96, indicating reasonable to very good agreement. Reg-Seq appears to be, if anything, more accurate than previous methods as it has higher relative information content in known areas of transcription factor binding and also does not have repressor-like bases on CRP sites as in Figure \ref{fig:compfig}(A) and (B).


\section{Extended details of analysis methods} \label{SI:analysis}
\subsection{Information footprints} \label{sec:Information}
    We use information footprints as a tool for hypothesis generation to identify regions which may contain transcription factor binding sites. In general, a mutation within a transcription factor site is likely to severely weaken that site. We look for groups of positions where mutation away from wild type has a large effect on gene expression. Our data sets consist of nucleotide sequences, the number of times we sequenced the construct in the plasmid library, and the number of times we sequenced its corresponding mRNA. A simplified data set on a 4 nucleotide sequence then might look like  \\

\begin{center}
    \begin{tabular}{c r r}
    Sequence & Library Sequencing Counts & mRNA Counts \\
    ACTA & 5 & 23\\
    ATTA & 5 & 3\\
    CCTG & 11 & 11\\
    TAGA & 12 & 3\\
    GTGC & 2 & 0\\
    CACA & 8 & 7\\
    AGGC & 7 & 3\\ 
    \label{table:fakedata}
    \end{tabular}
\end{center}

    One possible calculation to measure the impact of a given mutation on expression is to take all sequences which have base $b$ at position $i$ and determine the number of mRNAs produced per read in the sequencing library. By comparing the values for different bases we could determine how large of an effect mutation has on gene expression. However, in this paper we will use mutual information to quantify the effect of mutation, as \cite{Kinney2010} demonstrated could be done successfully. In Table 1 the frequency of the different nucleotides in the library at position 2 is 40$\%$ A, 32$\%$ C, 14$\%$ G and 14$\%$ T. Cytosine is enriched in the mRNA transcripts over the original library, as it now composes 68\% of all mRNA sequencing reads while A, G, and T only compose only 20$\%$, 6$\%$, and 6$\%$ respectively. Large enrichment of some bases over others occurs when base identity is important for gene expression. We can quantify how important using the mutual information between base identity and gene expression level. Mutual information is given at position $i$ by

\begin{equation}
    I_b =  \sum_{m=0}^1  \sum_{\mu=0}^1 p(m,\mu)\log_2\left(\frac{p(m,\mu)}{p_{mut}(m)p_{expr}(\mu)}\right).
    \label{eqn:MI2}
\end{equation}

 \noindent $p_{mut}(m)$ in equation ~\ref{eqn:MI2} refers to the probability that a given sequencing read will be from a mutated base. $p_{expr}(\mu)$ is a normalizing factor that gives the ratio of the number of DNA or mRNA sequencing counts to total number of counts. \\

    The mutual information quantifies how much a piece of knowledge reduces the entropy of a distribution. At a position where base identity matters little for expression level, there would be little difference in the frequency distributions for the library and mRNA transcripts. The entropy of the distribution would decrease only by a small amount when considering the two types of sequencing reads separately. \\

    We are interested in quantifying the degree to which mutation away from a wild type sequence affects expression. Although their are obviously 4 possible nucleotides, we can classify each base as either wild-type or mutated so that $b$ in equation \ref{eqn:MI2} represents only these two possibilities. \\

    If mutations at each position are not fully independent, then the information value calculated in equation \ref{eqn:MI2} will also encode the effect of mutation at correlated positions. If having a mutation at position 1 is highly favorable for gene expression and is also correlated with having a mutation at position 2, mutations at position 2 will also be enriched amongst the mRNA transcripts. Position 2 will appear to have high mutual information even if it has minimal effect on gene expression. Due to the DNA synthesis process used in library construction, mutation in one position can make mutation at other positions more likely by up to 10 percent. This is enough to cloud the signature of most transcription factors in an information footprint calculated using equation \ref{eqn:MI2}. \\

    We need to determine values for $p_i(m|\mu)$ when mutations are independent, and to do this we need to fit these quantities from our data. We assert that 

\begin{equation}
    \left\langle mRNA \right\rangle \propto e^{-\beta E_{eff}}
\end{equation}

    \noindent is a reasonable approximation to make. $\left\langle mRNA \right\rangle$ is the average number of mRNAs produced by that sequence for every cell containing the construct and $E_{eff}$ is an effective energy for the sequence that can be determined by summing contributions from each position in the sequence. There are many possible underlying regulatory architectures, but to demonstrate that our approach is reasonable let us first consider the simple case where there is only a RNAP site in the studied region. We can write down an expression for average gene expression per cell as

\begin{equation}
\left\langle mRNA \right\rangle \propto p_{bound} \propto \frac{\frac{p}{N_{NS}}e^{-\beta E_P}}{1 + \frac{p}{N_{NS}}e^{- \beta E_P}}
\label{eqn:pbound}
\end{equation}

    \noindent Where $p_{bound}$ is the probability that the RNAP is bound to DNA and is known to be proportional to gene expression in \textit{E. coli} ~\cite{garcia_quantitative_2011}, $E_P$ is the energy of RNAP binding, $N_{NS}$ is the number of nonspecific DNA binding sites, and $p$ is the number of RNAP. If RNAP binds weakly then $\frac{p}{N_{NS}}e^{-\beta E_P} << 1$. We can simplify equation ~\ref{eqn:pbound} to 

\begin{equation}
    \left\langle mRNA \right\rangle \propto e^{- \beta E_P}.
\end{equation}

    \noindent If we assume that the energy of RNAP binding will be a sum of contributions from each of the positions within its binding site then we can calculate the difference in gene expression between having a mutated base at position $i$ and having a wild type base as

\begin{align}
\frac{\left\langle mRNA_{WT_i} \right\rangle}{\left\langle mRNA_{Mut_i} \right\rangle} =& \frac{e^{- \beta E_{P_{WT_i}}}}{e^{- \beta E_{P_{Mut_i}}}} \\
\frac{\left\langle mRNA_{WT_i} \right\rangle}{\left\langle mRNA_{Mut_i} \right\rangle} =& e^{- \beta (E_{P_{WT_i}} - E_{P_{Mut_i}})}.
\end{align}

    \noindent In this example we are only considering single mutation in the sequence so we can further simplify the equation to

\begin{equation}
\frac{\left\langle mRNA_{WT_i} \right\rangle}{\left\langle mRNA_{Mut_i} \right\rangle} = e^{- \beta \Delta E_{P_i}}.
\end{equation}

    \noindent We can now calculate the base probabilities in the expressed sequences. If the probability of finding a wild type base at position $i$ in the DNA library is $p_i(m=WT|\mu=0)$ then 

\begin{align}
 p_i&(m = WT|\mu = 1) = \nonumber \\
&  \frac{p_i(m=WT|\mu=0) \frac{\left\langle mRNA_{WT_i} \right\rangle}{\left\langle mRNA_{Mut_i} \right\rangle}}{p_i(m=Mut|\mu=0)  + p_i(m=WT|\mu=0) \frac{\left\langle mRNA_{WT_i} \right\rangle}{\left\langle mRNA_{Mut} \right\rangle}} \\
p_i&(m=WT|\mu=1) =  \nonumber \\
& \frac{p_i(m=WT|\mu=0) e^{- \beta \Delta E_{P_i}}}{p_i(m=Mut|\mu=0)  + p_i(m=WT|\mu=0) e^{- \beta \Delta E_{P_i}}}.
\label{eqn:probeqn}
\end{align}

    Under certain conditions, we can also infer a value for $p_i(m|\mu=1)$ using a linear model when there are any number of activator or repressor binding sites. We will demonstrate this in the case of a single activator and a single repressor, although a similar analysis can be done when there are greater numbers of transcription factors. We will define $P = \frac{p}{N_{NS}}e^{- \beta E_P}$. We will also define $A = \frac{a}{N_{NS}}e^{-\beta E_A}$ where $a$ is the number of activators, and ${E_A}$ is the binding energy of the activator. We will finally define $R = \frac{r}{N_{NS}}e^{-\beta E_R}$ where $r$ is the number of repressors and ${E_R}$ is the binding energy of the repressor. We can write 
    
\begin{equation}
\left\langle mRNA \right\rangle \propto p_{bound} \propto \frac{P + PAe^{-\beta \epsilon_{AP}}}{1+A+P+R+PAe^{-\beta \epsilon_{AP}}}
\label{eqn:actrep}
\end{equation}

    If activators and RNAP bind weakly but interact strongly, and repressors bind very strongly, then we can simplify equation ~\ref{eqn:actrep}. In this case $A << 1$, $P << 1$, $PAe^{-\epsilon_{AP}} >> P$, and $R >> 1$. We can then rewrite equation ~\ref{eqn:actrep} as 

\begin{align}
\left\langle mRNA \right\rangle \propto & \frac{PAe^{-\beta \epsilon_{AP}}}{R} \\
\left\langle mRNA \right\rangle \propto & e^{-\beta(-E_P - E_A + E_R)} \label{eqn:infoenergy}
\end{align}

    \noindent As we typically assume that RNAP binding energy, activator binding energy, and repressor binding can all be represented as sums of contributions from their constituent bases, the combination of the energies can be written as a total effective energy $E_{eff}$ which is a sum of contributions from all positions within the binding sites. \\ 

     We fit the parameters for each base using a Markov Chain Monte Carlo Method. Two MCMC runs are conducted using randomly generated initial conditions. We require both chains to reach the same distribution to prove the convergence of the chains. We do not wish for mutation rate to affect the information values so we set the $p(WT) = p(Mut) = 0.5$ in the information calculation. The information values are smoothed by averaging with neighboring values. \\
 
 
\subsection{Analysis of mass spectrometry results}

Mass spectrometry results were processed using MaxQuant ~\cite{cox_maxquant_2008} ~\cite{cox_practical_2009}. Spectra were searched against the UniProt \textit{E. coli} K-12 database as well as a contaminant database (256 sequences). LysC was specified as the digestion enzyme. Proteins were considered if they were known to be transcription factors, or were predicted to bind DNA (using gene ontology term GO:0003677, for DNA-binding in BioCyc).

\subsection{Uncertainty due to number of independent sequences}
    1400 promoter variants were ordered from TWIST Bioscience for each promoter studied. Due to errors in synthesis, additional mutations are introduced into the ordered oligos. As a result, the final number of variants received was an average of 2200 per promoter. To test whether the number of promoter variants is a significant source of uncertainty in the experiment we computationally reduced the number of promoter variants used in the analysis of the \textit{zapAB} -10 RNAP region. Each sub-sampling was performed 3 times. The results, as displayed in Figure \ref{fig:downsample}, show that there is only a small effect on the resulting sequence logo until the library has been reduced to approximately 500 promoter variants. \\

\begin{figure}
\centering
\includegraphics[width=6.5truein]{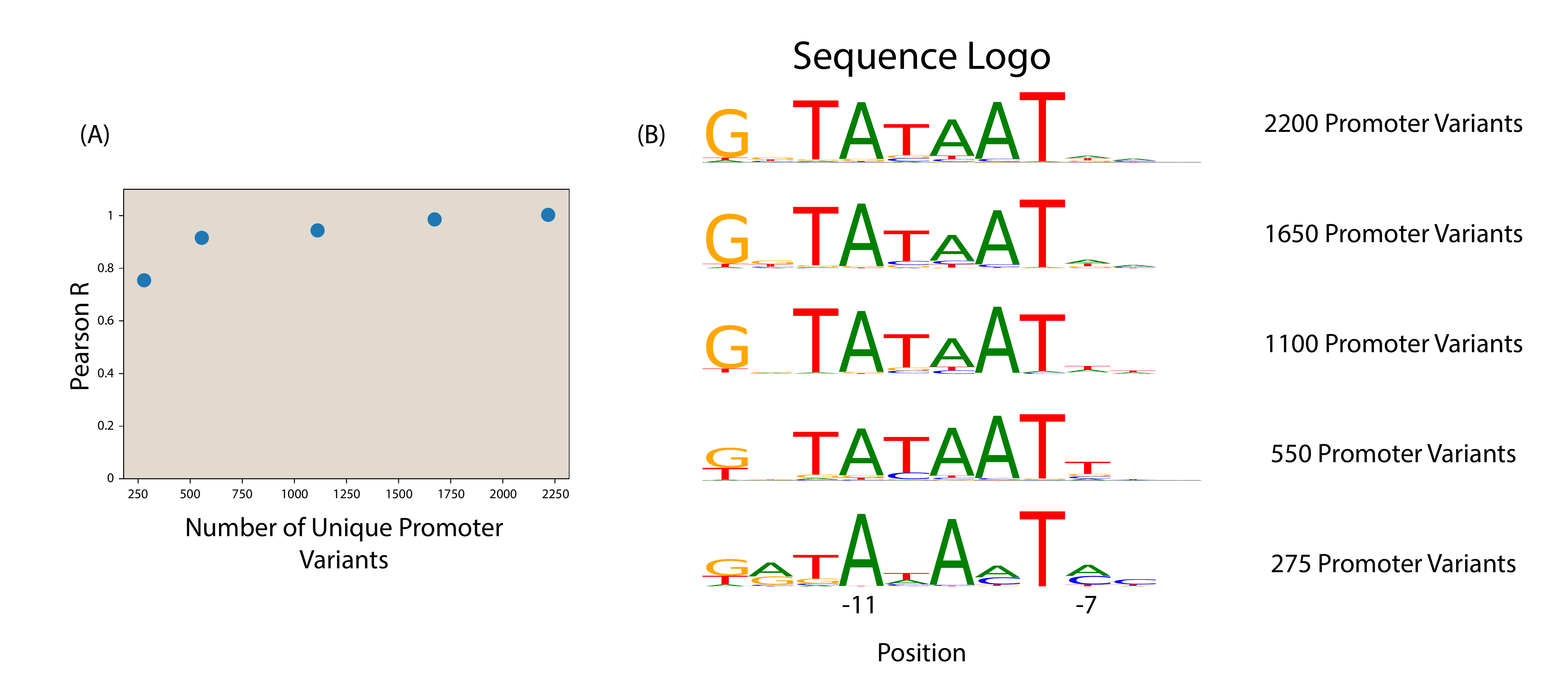}
\caption{A comparison of RNAP -10 site sequence logos. (A) This figure shows the Pearson correlation coefficient between the energy matrix models inferred from the full dataset (2200 unique promoter variants) and that from a computationally restricted dataset. (B) Sequence logos of the RNAP -10 region from each sub-sampled dataset.}
\label{fig:downsample}
\end{figure}

\subsection{TOMTOM motif comparison} \label{TOMTOM}
    In some cases, we used an alternative approach to mass spectrometry to discover the TF identity regulating a given promoter based on sequence analysis using a motif comparison tool. TOMTOM \cite{Gupta2007} is a tool that uses a statistical method to infer if a putative motif resembles any previously discovered motif in a database. Of interest, it accounts for all possible offsets between the motifs. Moreover, it uses a suite of metrics to compare between motifs such as Kullback-Leibler divergence, Pearson correlation, euclidean distance, among others. \\
    
    We performed comparisons of the motifs generated from our energy matrices to those generated from all known transcription factor binding sites in RegulonDB. Figure \ref{fig:tomtom} shows a result of TOMTOM, where we compared the motif derived from the -35 region of the \textit{ybjX} promoter and found a good match with the motif of PhoP from RegulonDB. \\
    
    The information derived from this approach was then used to guide some of the TF knockout experiments, in order to validate its interaction with a target promoter characterized by the loss of the information footprint. Furthermore, we also used TOMTOM to search for similarities between our own database of motifs, in order to generate regulatory hypotheses in tandem. This was particularly useful when looking at the group of GlpR binding sites found in this experiment. 
    
\begin{figure}
\centering
\includegraphics[width=5.0truein]{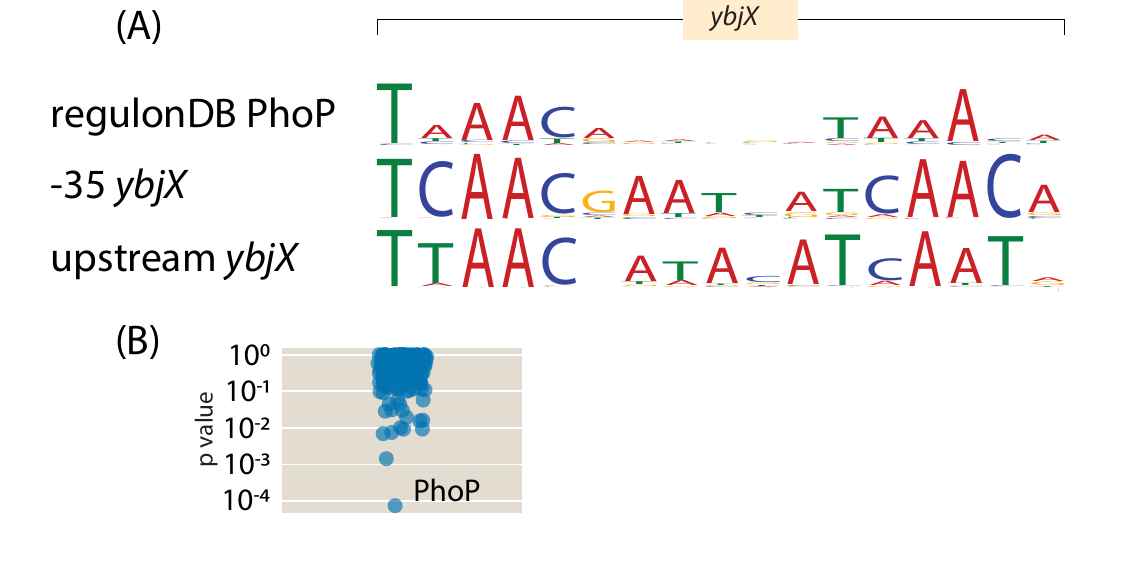}
\caption{Motif comparison using TOMTOM. Searching our energy motifs against the RegulonDB database using TOMTOM allowed us to guide our TF knockout experiments. Here we show the sequence logos of the PhoP transcription factor from RegulonDB (top) and the one generated from the ybjX promoter energy matrix. E-value = 0.01 using Euclidean distance as a similarity matrix.}
\label{fig:tomtom}
\end{figure}

\section{Additional results} \label{SI:results}

\subsection{Binding sites regulating divergent operons}

    In addition to discovering new binding sites, we have discovered additional functions of known binding sites. In particular, in the case of \textit{bdcR}, the repressor for the divergently  transcribed gene \textit{bdcA} \cite{partridge_nsrr_2009-1}, is also shown to repress \textit{bdcR} in Figure \ref{fig:diverge}(A). Similarly in Figure \ref{fig:diverge}(B) IvlY is shown to repress \textit{ilvC} in the absence of inducer. Divergently transcribed operons that share regulatory regions are plentiful in \textit{E. coli}, and although there are already many known examples of transcription factor binding sites regulating several different operons, there are almost certainly many examples of this type of transcription that have yet to be discovered. \\
    
\begin{figure}
\centering
\includegraphics[width=6.5truein]{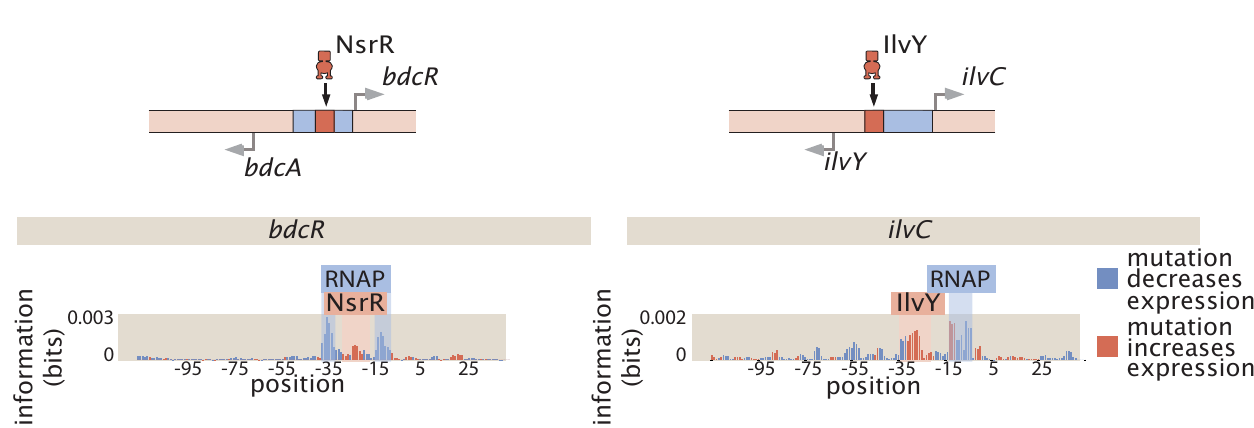}
\caption{Two cases in which we see transcription factor binding sites that we have found to regulate both of the two divergently transcribed genes.}
\label{fig:diverge}
\end{figure}
        
    Multi-purpose binding sites allow for more genes to be regulated with fewer binding sites. However, they can also serve to sharpen the promoter's response to environmental cues. In the case of \textit{ilvC}, IlvY is known to activate \textit{ilvC} in the presence of inducer. However, we now see that it also represses the promoter in the absence of that inducer. The production of \textit{ilvC} is known to increase by approximately a factor of 100 in the presence of inducer \cite{rhee_activation_1998}. The magnitude of the change is attributed to the cooperative binding of two IlvY binding sites, but the lowered expression of the promoter due to IlvY repression in the absence of inducer is also a factor. 

\subsection{Regulatory cartoons}

\begin{figure}[!h]
\centering
\includegraphics[width=7.0truein]{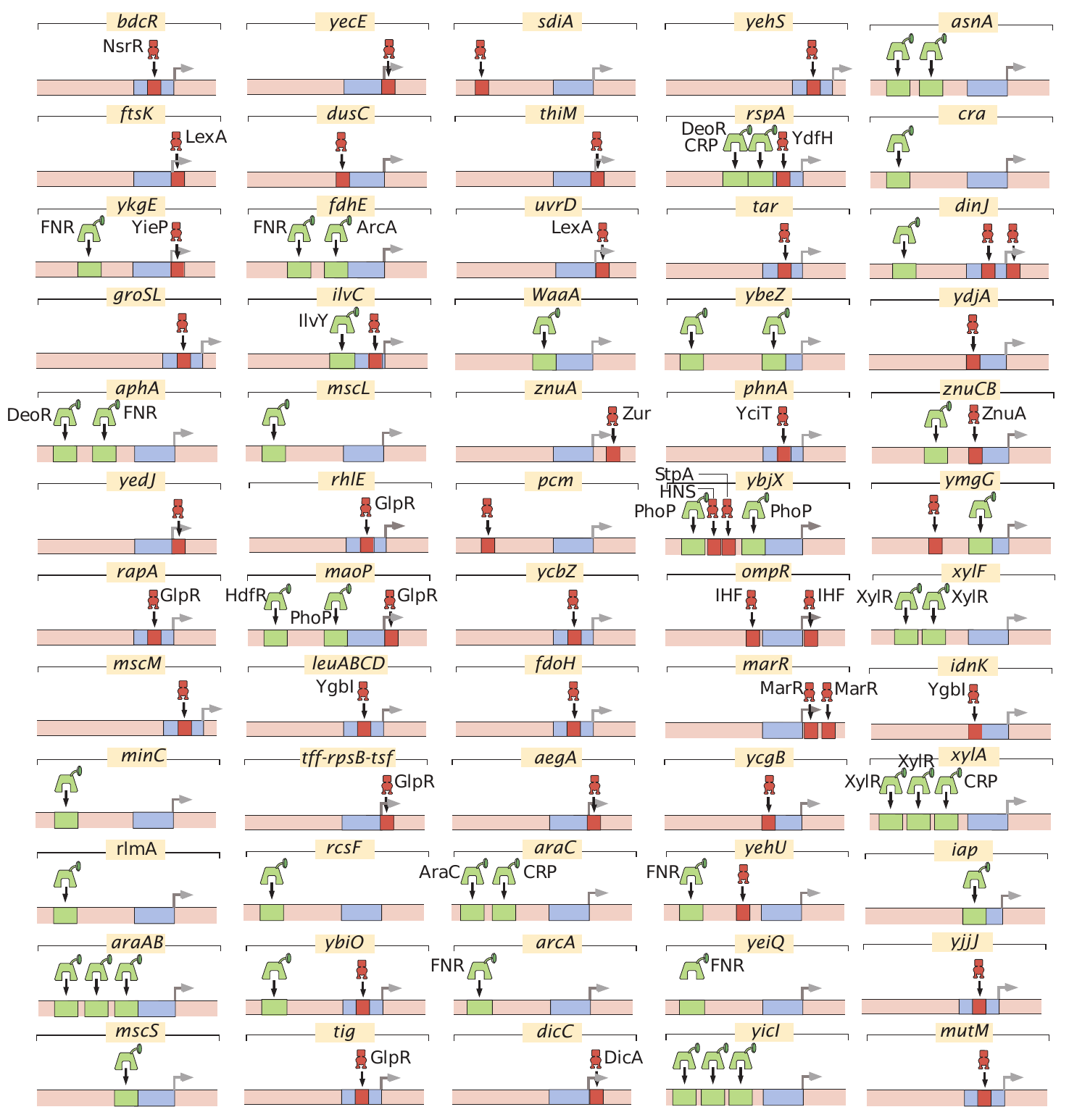}
\caption{All regulatory cartoons for genes considered in our study.}
\label{fig:cartoons}
\end{figure}

\pagebreak

\subsection{Comparison of results to regulonDB}

\begin{figure}[!h]
\centering
\includegraphics[width=5.0truein]{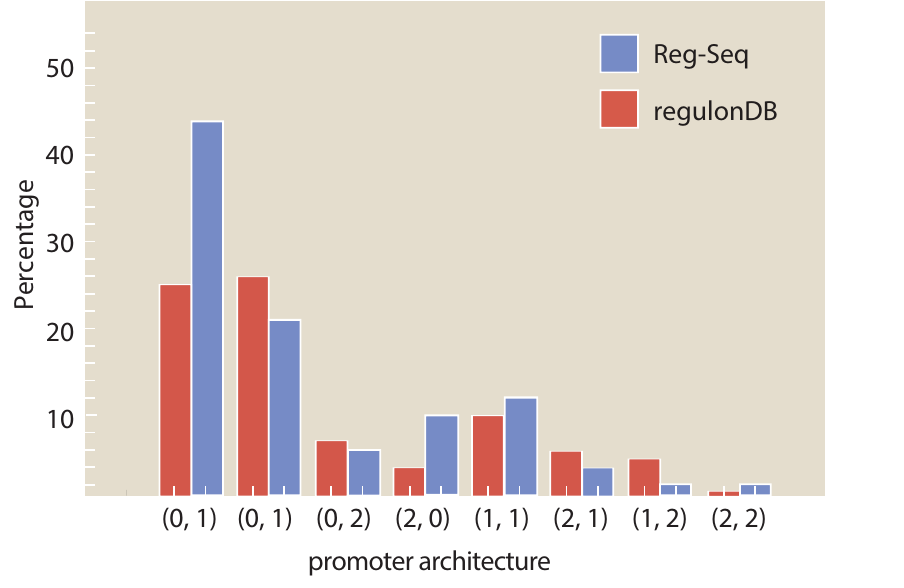}
\caption{A comparison of the types of architectures found in RegulonDB \cite{Santos_Zavaleta2019} to the architectures with newly discovered binding sites found in the Reg-Seq study.}
\label{fig:comp_to_DB}
\end{figure}

\end{document}